\PassOptionsToPackage{table}{xcolor}
\documentclass{comsoc2023}

\usepackage[hidelinks]{hyperref}
\usepackage{amsfonts}
\usepackage{amsthm}
\usepackage{algorithm}
\usepackage{mathtools}
\usepackage{pifont}
\usepackage{bbm}
\usepackage{algpseudocode}
\usepackage{comment}
\usepackage[table]{xcolor}
\usepackage{placeins}
\usepackage{pbox}
\usepackage{cleveref}
\usepackage{wrapfig}
\usepackage[bibliography=common]{apxproof}
\usepackage{chngcntr}
\counterwithin{equation}{section}

\usepackage{comment}
\usepackage{placeins}
\usepackage{tikz}
\usetikzlibrary{decorations.text}

\newtheorem{definition}{Definition}[section]
\newtheorem{example}{Example}
\newtheorem{theorem}{Theorem}
\newtheorem{lemma}[theorem]{Lemma}

\newtheoremrep{theorem}{Theorem}
\newtheoremrep{lemma}{Lemma}

\newcommand{\cmark}{\ding{51}}%
\newcommand{\xmark}{\ding{55}}%
\DeclarePairedDelimiter\ceil{\lceil}{\rceil}
\DeclarePairedDelimiter\floor{\lfloor}{\rfloor}

\DeclareMathOperator*{\argmax}{arg\,max}

\DeclarePairedDelimiterX\set[1]\lbrace\rbrace{#1}

\title{The Leximin Approach for a Sequence of Collective Decisions}

\author{Ido Kahana and Noam Hazon
}

\begin{document}


\begin{abstract}
In many situations, several agents need to make a sequence of decisions. For example, a group of workers that needs to decide where their weekly meeting should take place. In such situations, a decision-making mechanism must consider fairness notions. In this paper, we analyze the fairness of three known mechanisms: round-robin, maximum Nash welfare, and leximin. We consider both offline and online settings, and concentrate on the fairness notion of proportionality and its relaxations. Specifically, in the offline setting, we show that the three mechanisms fail to find a proportional or approximate-proportional outcome, even if such an outcome exists. We thus introduce a new fairness property that captures this requirement, and show that a variant of the leximin mechanism satisfies the new fairness property. In the online setting, we show that it is impossible to guarantee proportionality or its relaxations. We thus consider a natural restriction on the agents' preferences, and show that the leximin mechanism guarantees the best possible additive approximation to proportionality and satisfies all the relaxations of proportionality.
\end{abstract}


\section{Introduction}
\label{Chapter1}

Commonly, a group of agents needs to reach a collective decision, thus needing a collective decision-making mechanism. For example, friends choosing a restaurant may utilize a voting procedure for deciding where to go. Several voting procedures, as well as other collective decision-making mechanisms, have been developed for reaching a one-time, single decision, but there are many situations in which there is a sequence of decisions. For example, consider a city council that needs to decide, each month, which activity to subside. It can choose to organize a family activity in the public park, hold a concert in the community center, or operate an overnight bus, but only one activity can be subsided each month. A senior citizen may benefit the most from a concert, but she may also benefit, to a lesser extent, from the overnight bus. A parent with young kids may benefit the most from an activity in the park, but she may also benefit from a concert. In summary, every activity benefits all the agents, but they may evaluate the activities differently.
Clearly, the city council would like to choose activities that satisfy all of the citizens, and it may take advantage of the fact that the decision situation is repeated every month.
As another example, consider friends that study together for an exam, and they need to choose a restaurant every day. Clearly, if most of them prefer a pizza, then it is reasonable that they will go together to a pizzeria on many days, even though one of the friends, say Bob, prefers sushi. However, since there is a sequence of decisions, it is also reasonable to consider fairness, i.e., an outcome in which the group of friends goes to a pizzeria every day may not be fair for Bob.

Arguably, one of the most fundamental notions of fairness is proportionality ($Prop$). 
That is, assume there is a sequence of decisions, and we call each decision situation a round. There are $n$ agents, and each agent evaluates the candidates in each round. We would like that each one of the agents will get at least a $1/n$ fraction of the utility she would get if she could solely decide on the outcome in each round.
Unfortunately, there are instances in which a proportional outcome does not exist, and thus it is reasonable to consider a multiplicative approximation of proportionality, namely, $\alpha$-$Prop$. Alternatively, \citeauthor{conitzer2017fair}~\cite{conitzer2017fair} have suggested two relaxations of proportionality, namely, $Prop1$ and $RRS$ (see ~\autoref{sec:definition} for the formal definitions).

When dealing with a sequence of decisions, it is important to distinguish between offline and online settings. In the offline setting, the decision-making mechanism gets as input all the valuations of the agents in all the rounds, and it needs to choose an outcome for every round. In the online setting, at each round, the mechanism gets as an input only the information up to this round, and it thus needs to choose one outcome (for this round). Unfortunately, fairness considerations have been largely neglected in the online setting,  and only a few works have analyzed fairness in such a setting. Indeed, it is hard to guarantee fairness in the online setting if the valuations of the agents are unrestricted. One possible restriction to the agents' valuations is based on the Borda scoring rule. Specifically, if there are $m$ candidates, it is assumed that, for each agent, the valuation of her most preferred candidate is $m-1$, the valuation of her second preferred candidate is $m-2$, and so on; the valuation of her least preferred candidate is $0$. 
We denote such valuations as Borda valuations. Indeed, the restriction to Borda valuations is also useful in the offline setting, for translating ordinal preference orders to cardinal preferences \cite{lu2011budgeted,baumeister2014scoring,darmann2015maximizing,darmann2016proportional}. That is, there are settings in which it is easier for the agents to express their preferences using ordinal preference orders, in which each agent reports a total order over the set of candidates (i.e., a ranking) for every round. In these settings, it was suggested to translate any ordinal preference order to numerical values according to the Borda scoring rule: the valuation of the highest-ranked candidate is $m-1$, the valuation of the second-ranked candidate is $m-2$, and so on.

In this paper, we study collective decision-making mechanisms for a sequence of decisions, both in the offline and online setting, and we analyze three common mechanisms: round-robin ($RR$), maximum Nash welfare ($MNW$), and leximin ($LMin$).
We first claim that an outcome that satisfies the relaxations of proportionality that were previously suggested (i.e., $Prop1$ and $RRS$) might be ``far'' from being proportional. Specifically, in the offline setting, we show that even though the $RR$ and $MNW$ mechanisms satisfy $Prop1$ and the $LMin$ mechanism satisfies $RRS$ (with an $RRS$-based normalization), the three mechanisms might fail to find a proportional outcome, even if such an outcome exists.
We thus introduce a new natural fairness property, Max-Possible-Prop ($MPP$). According to this property, a mechanism should return an outcome that is as proportional as possible. For example, if an instance admits an outcome that satisfies $Prop$, then the mechanism should return an outcome that satisfies $Prop$. In addition, if an instance does not admit an outcome that satisfies $Prop$ but admits an outcome that satisfies $1/2$-$Prop$, then the mechanism should return an outcome that satisfies $1/2$-$Prop$. Generally, a mechanism that satisfies $MPP$ returns an outcome that is $\alpha$-proportional, with the maximum possible $\alpha$ for the given instance. 

We show that there is an unavoidable trade-off between $MPP$ and $Prop1$, and between $MPP$ and $RRS$ when there are at least $3$ agents. That is, a mechanism that satisfies $MPP$ does not guarantee any constant factor approximation of $RRS$ or $Prop1$. However, we show that a leximin mechanism, in which the valuations are normalized with the proportional value of each agent (i.e., a $Prop$-based normalization), satisfies $MPP$. 
Moreover, with two agents, this mechanism satisfies $MPP$, $RRS$, and $1/2$-$Prop1$.

We then analyze the restricted setting of Borda valuations.
In this setting, an outcome that satisfies $Prop1$ is ``far'' from being proportional by an additive factor of at most $m-1$, and we thus consider an additive approximation of $Prop$.
We show that, unfortunately, both $RR$ and $MNW$ do not guarantee an additive constant-factor approximation of $Prop$. On the other hand, the leximin mechanism satisfies $MPP$, $RRS$, $Prop1$, and it guarantees a $1$-additive approximation of $Prop$, which is the best possible constant factor additive approximation of $Prop$. 
We note that \citeauthor{conitzer2017fair}~\cite{conitzer2017fair} have raised an open question: 
is there a mechanism that satisfies $PO$, $Prop1$ and $RRS$ simultaneously? We partially solve this question- if we restrict the valuations to Borda valuations, then $LMin$ is such a mechanism, since we show that it satisfies $PO$, $Prop1$, and $RRS$.

In the online setting, we show that it is impossible to achieve proportionality and even the weaker fairness properties (i.e., $Prop1$, $RRS$, and $MPP$). We thus consider the restriction to Borda valuations, and show that the online leximin mechanism guarantees a $1$-additive approximation of $Prop$. Moreover, the online leximin mechanism satisfies $Prop1$ and $RRS$. However, we show that there is no online mechanism that satisfies $MPP$, even with Borda valuations.

The main contributions of this paper are threefold. We introduce a natural fairness property, which is (arguably) a better relaxation of proportionality than $RRS$ and $Prop1$, and show that a variant of the offline leximin mechanism satisfies it. We also partially solve an open question of \citeauthor{conitzer2017fair}~\cite{conitzer2017fair}, by using the restriction to Borda valuation. Finally, in the online setting with Borda valuations, we show that the leximin mechanism guarantees the best possible constant factor additive approximation of $Prop$. 

Tables \ref{tab:offline}, and \ref{tab:borda} summarize our results. Note that the full proofs of some of the theorems are deferred to the appendix due to space constraints.

\begin{table}[ht]
    \footnotesize
    \centering
\setlength{\extrarowheight}{5pt}
\setlength\arrayrulewidth{2pt}
\begin{tabular}{@{\extracolsep{\fill}} |c|c|c|c|c|c|c|}
 \hline
  & $PO$ & $Prop$ & $Prop1$ & $MPP$ & $RRS$ \\
 \hline
  $\textsf{LMin}^{\textsf{Prop}}_{\textsf{off}}$ n = 2 & \cmark & {\xmark} & {1/2} & \cmark & {\cmark} \\
   \hline
  $\textsf{LMin}^{\textsf{Prop}}_{\textsf{off}}$ $n > 2$   & \cmark & {\xmark} & {\xmark} & {\cmark} & {\xmark}  \\
 \hline
  $\textsf{LMin}^{\textsf{RRS}}_{\textsf{off}}$ & \cellcolor{gray!25}\cmark & \cellcolor{gray!25}{\xmark} & \cellcolor{gray!25}1/2  & \xmark & \cellcolor{gray!25}\cmark \\
  \hline
  $\textsf{RR}_{\textsf{off}}$           & \cellcolor{gray!25}\xmark & \cellcolor{gray!25}{\xmark} & \cellcolor{gray!25}\cmark & \xmark & \cellcolor{gray!25}\cmark  \\
  \hline
$\textsf{MNW}_{\textsf{off}}$             & \cellcolor{gray!25}\cmark &  \cellcolor{gray!25}{\xmark} &\cellcolor{gray!25} \cmark  & \xmark & \cellcolor{gray!25} 1/n \\
 \hline
\end{tabular}
 \caption{Summary of results for the offline setting, where there are no restrictions on the valuation. The results in gray are due to \protect\cite{conitzer2017fair}.}
 \label{tab:offline}
\end{table}

\begin{table}[ht]
    \footnotesize
    \centering
\setlength{\extrarowheight}{5pt}
\setlength\arrayrulewidth{2pt}
\begin{tabular}{@{\extracolsep{\fill}} |c|c|c|c|c|c|c|}
 \hline
 & $PO$ & $Prop$ & $Prop1$ & $MPP$ & $RRS$\\
 
 \hline
   $\textsf{LMin}_{\textsf{off}}$       &   \cellcolor{gray!25}\cmark & $1$-additive  & \cmark  & \cmark & \cellcolor{gray!25}\cmark \\
  \hline
 $\textsf{RR}_{\textsf{off}}$            & \xmark & $m$-$1$-additive  & \cellcolor{gray!25}\cmark & \xmark & \cellcolor{gray!25}\cmark  \\
  \hline
$\textsf{MNW}_{\textsf{off}}$               & \cellcolor{gray!25}\cmark &  \vtop{\hbox{\strut $x$-additive}\hbox{\strut $m$-$1 \geq$ x}\hbox{\strut x $\geq \frac{m-3}{2}$}} & \cellcolor{gray!25}\cmark & \xmark & \xmark \\
 \hline
  $\textsf{LMin}_{\textsf{on}}$       & \xmark & $1$-additive   & \cmark  & \xmark & \cmark \\
  \hline
 $\textsf{RR}_{\textsf{on}}$             & \xmark & $m$-$1$-additive  & \cmark & \xmark  & \cmark  \\
  \hline
$\textsf{MNW}_{\textsf{on}}$              & \xmark & 

\vtop{\hbox{\strut $x$-additive}\hbox{\strut x $\geq \frac{m-3}{2}$}} & ? & \xmark & \xmark \\
 \hline
\end{tabular}
 \caption{ Summary of results with Borda valuations. The results in gray are due to \protect\cite{conitzer2017fair}.}
 \label{tab:borda}
\end{table}

\section{Related Works} 
\label{Chapter2}
The analysis of collective decision-making mechanisms for a sequence of decisions has been studied both in the domain of fair division, in which it is commonly called public decision-making \cite{conitzer2017fair}, and in the domain of voting, in which it is commonly called perpetual voting \cite{lackner2020perpetual}. 
Specifically, \citeauthor{conitzer2017fair}~\cite{conitzer2017fair} introduce the problem of public decision-making in the offline setting. They propose $RRS$ and $Prop1$ as relaxations of proportionality, and provide a comprehensive analysis of the $RR$, $MNW$, and $LMin$ mechanisms. We note that their analysis of the  $LMin$ mechanism uses an $RRS$-based normalization, while we propose a $Prop$-based normalization. In addition, we provide an alternative relaxation of proportionality, $MPP$, analyze the restricted setting of Borda valuations, and the online setting. \citeauthor{freeman2017fair}~\cite{freeman2017fair} study the online version of public decision-making. They concentrate on maximizing the Nash welfare, and present two greedy mechanisms. However, they do not analyze their mechanisms with regard to proportionality or its relaxations.

In the domain of voting, \citeauthor{lackner2020perpetual}~\cite{lackner2020perpetual} suggests several online voting rules when there is a sequence of decisions, and analyzes them via three axiomatic properties, as well as a quantitative evaluation by computer simulations. In a subsequent paper, \citeauthor{lackner23prop}~\cite{lackner23prop} define two classes of voting rules that are particularly easy to explain to voters, and define specific proportionality axioms. \citeauthor{bulteau2021justified}~\cite{bulteau2021justified} study the offline setting, and analyze the fairness for subgroups of voters by adapting the well-established Justified Representation (JR) and Proportional Justified Representation (PJR) axioms. \citeauthor{skowron2022proportional}~\cite{skowron2022proportional} also study the offline setting, and propose a proportionality notion that ensures guarantees to all groups of voters. 
All of these works assume that the voters express approval preferences, while we study cardinal preferences (or ordinal preferences that are translated to cardinal preferences with the Borda scoring rule). 

Our model is closely related to other frameworks that have been studied in computational social choice. Specifically, the framework of multi-winner voting \cite{faliszewski2017multiwinner,lackner2023multi} considers fairness properties, and the outcome consists of several winning candidates, as in our setting. However, in multi-winner voting, a candidate cannot be elected multiple times, as in our setting. Notably, \citeauthor{votes2022change}~\cite{votes2022change} study a sequence of multi-winner elections, in which  the difference between the winners in consecutive rounds is upper-bounded. The framework of participatory budgeting \cite{aziz2021participatory,rey2023computational}, which generalizes multi-winner voting, utilizes voting systems  for deciding on the funding of public projects. \citeauthor{lackner2021fairness}~\cite{lackner2021fairness} study a sequence of participatory budgeting problems, and introduce a theory of fairness for this setting. 
Our setting can also be applied for modeling a sequence of voting on the funding of public projects, but there is no budget constraint.
The model of fair allocation of indivisible public goods \cite{fain2018fair} generalizes participatory budgeting as well as our model of public decision-making. In this model, there is a set of public goods and feasibility constraints on what subsets of goods can be chosen.  
In the offline setting, \citeauthor{fain2018fair}~\cite{fain2018fair} provide an additive approximation to the core, which is a fairness notion for groups of agents. \citeauthor{garg2021fair}~\cite{garg2021fair} consider the simple constraint in which the number of public goods is bounded, and analyze the maximum Nash welfare and leximin objectives with regard to the $RRS$ and $Prop1$ fairness properties. \citeauthor{banerjee2022proportionally}~\cite{banerjee2022proportionally}  study an online version of fair allocation of public goods, and consider proportional fairness.

\section{Definitions}
\label{sec:definition}
Consider a set of agents $N = \set{1,2,\ldots,n}$, and a set of candidates $C = \set{c_1,c_2,\ldots,c_m}$ \footnote{For the clarity of presentation, we assume that the set of candidates is static. All of our results are easily adapted to the setting in which the set of candidates changes from round to round.
}.
For every round $t=1, \ldots, T$, every agent $i$ reports her valuation $v^t_i(c_j) \in \mathbb{R}_{\geq 0}$ for every candidate $c_j$. 
We assume that every agent $i$ has at least one positive valuation.
For a given round $t$, we can write the valuations of all the agents in a matrix, denoted by $V^t$, where $V^t = {(\boldsymbol{v}^t_i(c_j))}_{ij}$. 
We denote by $\boldsymbol{v^t}(c_j)$ the vector of valuations that all the agents assign to candidate $c_j$.
We investigate mechanisms that choose an outcome $\boldsymbol{o}=(o^1,\ldots,o^T)$, $o^t \in C$, which is a choice of a candidate for every round.
We assume that the agents have additive utility functions. Therefore, we define the accumulated utility of an agent $i$ from outcome $\boldsymbol{o}$, denoted by $u_i(\boldsymbol{o})$, as  $u_i(\boldsymbol{o}) = \sum_{t = 1}^{T} v^t_i(o^t)$. The accumulated utility vector of all the agents, denoted by $\boldsymbol{u}(\boldsymbol{o})$, is a vector in which the $i$-th entry is $u_i(\boldsymbol{o})$.

In the offline setting, the mechanism gets as input all of the agents' valuations in all the rounds, i.e., it gets the vector $(V^1,\ldots,V^T)$, and chooses the outcome $\boldsymbol{o}$.
In the online setting, the mechanism determines the outcome $\boldsymbol{o}$ sequentially, i.e., the mechanism determines each $o^t$ at round $t$, since the mechanism does not get the entire input upfront. Indeed, at each round $t$, the mechanism gets as an input only the information up to this round, i.e., the vector $(V^1,\ldots,V^t)$, and the vector of chosen candidates  $\boldsymbol{o}^{t-1} = (o^1,\ldots,o^{t-1})$.
We slightly abuse notation and define the accumulated utility of agent $i$ up to round $t$ from outcome $o^{t-1}$ as $u_i(\boldsymbol{o}^{t-1}) = \sum_{k = 1}^{t-1} v^k_i(o^k)$. Similarly, the accumulated utility vector of all the agents up to round $t$ is $u(\boldsymbol{o}^{t-1})$. Clearly, at the first round, the utility of every agent $i$ is zero (i.e., $u_i(\boldsymbol{o}^0) = 0$).

There are settings in which it is easier for the agents to express their preferences using ordinal preference orders. That is, each agent reports for every round a total order over $C$. In these settings, we translate the ordinal preference orders to cardinal preferences with the Borda scoring rule. That is, for each agent, the valuation of a candidate $c$ is the number of candidates ranked below $c$. We denote such valuations as Borda valuations.

\section{Efficiency and Fairness}
\label{subsec:eff_fair}
When analyzing offline or online mechanisms, we focus on popular notions of efficiency and fairness. 
Specifically, for efficiency, we consider the notion of Pareto optimality, which is defined as follows:
\begin{definition}
\label{defn:po}
An outcome $\boldsymbol{o}$ is \textit{Pareto Optimal} (PO) if there does not exist another outcome $\boldsymbol{o}'$ such that for every agent $i$, (1) $u_i(\boldsymbol{o}') \geq u_i(\boldsymbol{o})$ and (2) there exists an agent $j$ such that $u_j(\boldsymbol{o}') > u_j(\boldsymbol{o})$.
\end{definition}
A mechanism satisfies $PO$ if it always chooses an outcome that is $PO$. Generally, we say that a mechanism satisfies a given efficiency or fairness property if it always chooses an outcome that satisfies this property.

For fairness, we concentrate on proportionality, which requires that each agent will receive at least $\frac{1}{n}$ fraction of the utility she would receive had she chosen the outcome.
Given an agent $i$, let ${cMax}^t_i$ be a candidate with the highest valuation at round $t$, i.e., ${cMax}^t_i \in \argmax_{c \in C} v^t_i(c)$. 
 
\begin{definition}
Let $Prop_i = \frac{1}{n}\sum_{t=1}^T v^t_i({cMax}^t_i).$ For $\alpha \in (0,1]$, we say that an outcome $\boldsymbol{o}$ satisfies $\alpha$-proportionality ($\alpha$-$Prop$) if for every agent $i$, $u_i(\boldsymbol{o}) \geq \alpha \cdot Prop_i$.
\end{definition}
We denote $1$-$Prop$ by $Prop$.
Unfortunately, there are instances in which there is no outcome that satisfies $\alpha$-$Prop$, for any constant $\alpha$ (i.e., an $\alpha$ that does not depend on the given instance). Therefore, \citeauthor{conitzer2017fair}~\cite{conitzer2017fair} propose a relaxation of $\alpha$-$Prop$, which is proportionality up to one round.
\begin{definition}
An outcome $\boldsymbol{o}$ satisfies $\alpha$-$Prop1$ if for every agent $i \in N$ there exists a round $t$ such that changing $o^t$ to ${cMax}^t_i$ ensures that agent $i$ receives a utility of at least $\alpha$-$Prop_i$,  i.e., 
$\forall_{i \in N} \exists_{t}, u_i(\boldsymbol{o}) - v^t_i(o^t) + v^t_i({cMax}^t_i) \geq \alpha \cdot Prop_i .$
\end{definition}
We denote $1$-$Prop1$ by $Prop1$.
Indeed, a mechanism may satisfy $\alpha$-$Prop1$, but it may still choose an outcome that does not satisfy $\alpha$-$Prop$, even though such an outcome exists. We thus propose a new fairness property, Max-Possible-Prop ($MPP$).
\begin{definition}
An outcome $\boldsymbol{o}$ satisfies Max-Possible-Prop ($MPP$) if for each $\alpha$ with which there is an outcome that satisfies $\alpha$-$Prop$,  $\boldsymbol{o}$ also satisfies $\alpha$-$Prop$.  
\end{definition}
That is, an outcome that satisfies the $MPP$ property satisfies $\alpha$-$Prop$ with the maximum possible $\alpha$ for the given instance.

We also propose to consider an additive approximation of $Prop$. Formally, an outcome $\boldsymbol{o}$ satisfies a $\beta$-additive approximation of $Prop$ if for every agent $i$, $u_i(\boldsymbol{o}) + \beta \geq Prop_i$, for $\beta > 0$.

Another notion of fairness is round-robin share (RRS) \cite{conitzer2017fair}.
For an agent $i$, let $\boldsymbol{cMax_i}$ be a vector that contains all the values $v^t_i({cMax}^t_i)$, $1 \leq t \leq T$, when they are sorted in a non-ascending order. We denote by $\boldsymbol{cMax_i}(t)$ the element in the $t$-th entry of $\boldsymbol{cMax_i}$.

\begin{definition}
Let $RRS_i = \sum_{1 \leq t \leq \floor{T/n}}\boldsymbol{cMax_i}(t \cdot n).$ For $\alpha \in (0,1]$
we say that an outcome $\boldsymbol{o}$ satisfy $\alpha$-round-robin share ($\alpha$-RRS) if for every agent $i$, $u_i(\boldsymbol{o}) \geq  \alpha \cdot RRS_i$. 
\end{definition}
We denote $1$-$RRS$ by $RRS$.

\section{Mechanisms}
We concentrate on three families of mechanisms, where each family consists of offline and online mechanisms.

\paragraph{Round robin (RR)}

In this mechanism there is a given order over the agents, and they take turns according to this order. In the offline RR, when an agent's turn arrives she chooses a round (that has not been chosen yet), and determines the winning outcome for this round. It is assumed that the agent will choose a round that yields her the highest utility \cite{conitzer2017fair}.
The online RR chooses a single outcome at each round $t$, and the order over the agent associates the rounds with the agents. Thus, when an agent's turn arrives, she only determines the winning outcome for the associated round $t$.
Formally, let $\pi$ be a permutation over $N$.
The offline RR, denoted by $\textsf{RR}_{\textsf{off}}$, chooses an outcome $\boldsymbol{o}=(o^1,\ldots,o^T)$ according to ~\autoref{alg:RR_offline}.

\begin{algorithm}[H]
   \caption{RR offline}
   \label{alg:RR_offline}
    \begin{algorithmic}[1] 
        \State $rounds \gets \{1,\ldots,T\}$
        \State $k \gets 1$
        \While{$|rounds| > 0$}
        \State $i \gets \pi(1+ (k-1) \pmod n)$
        \State $t \gets \argmax_{r \in rounds} v^r_i({cMax}_{i}^{r})$
        \State $o^{t} \gets {cMax}_{i}^{t}$
        \State $rounds \gets rounds \setminus \{t\}$
        \State $k \gets k + 1$
        \EndWhile
    \end{algorithmic}
\end{algorithm}

Given a permutation $\pi$, let $turn^t(\pi)$ be a function that associates a round $t$ with an agent, according to the order $\pi$, $turn^t(\pi) = \pi(1+ (t-1) \pmod n)$.
The online RR is defined as follows.
\begin{definition}
Given a permutation $\pi$ over $N$, the online RR ($\textsf{RR}_{\textsf{on}}$) chooses an outcome $o^t$ for round $t$ such that $o^t = cMax_{turn^t(\pi)}^t$.
\end{definition}


\paragraph{Maximum Nash welfare (MNW)}
The Nash welfare of an outcome is the product of the utilities of all the agents. The offline MNW mechanism ($\textsf{MNW}_{\textsf{off}}$) chooses an outcome that maximizes the Nash welfare. The online MNW mechanism ($\textsf{MNW}_{\textsf{on}}$) chooses an outcome $o^t$ for round $t$ that maximizes the Nash welfare up to round $t$. Formally, 
\begin{definition}
$\textsf{MNW}_{\textsf{off}}$ chooses an outcome $\boldsymbol{o}$ such that $\boldsymbol{o} \in \argmax_{\boldsymbol{o'}} \prod_{i \in N} u_i(\boldsymbol{o'}).$
\end{definition}
\begin{definition}
$\textsf{MNW}_{\textsf{on}}$ chooses an outcome $o^t$ for round $t$ such that $o^t \in \argmax_{c_j \in C} \prod_{i \in N} (u_i(\boldsymbol{o}^{t-1}) + v^t_i(c_j)).$
\end{definition}
When all outcomes have a Nash welfare of $0$, $\textsf{MNW}_{\textsf{off}}$ and $\textsf{MNW}_{\textsf{on}}$ find the largest set of agents that there is an outcome that gives them positive utilities, and choose an outcome that maximizes the product of the agents' utilities.
Note that both RR and MNW mechanisms are scale-free. That is, the units of measurement used by the agents for expressing their valuations do not affect the outcome.

\paragraph{Leximin}
Generally, the motivation behind the mechanisms in this family is to maximize the utility of the agent that has the minimum utility, i.e., the worst-off agent. However, since there might be several such outcomes, a leximin mechanism chooses an outcome that also maximizes the utility of the second worst-off agent, and then the third, and so forth. This idea is formalized by the leximin ordering.
Given a vector $\boldsymbol{u}$, let $\boldsymbol{\widetilde{u}}$ be the vector $\boldsymbol{u}$ when the elements are ordered in a non-descending order.
\begin{definition}
The leximin ordering, $\succ$, is a total preorder, in which for any two vectors with the same number of elements, $\boldsymbol{x}, \boldsymbol{y}$,
$\boldsymbol{x} \succ \boldsymbol{y}$ if there exists an index $i$ such that $\boldsymbol{\widetilde{x}}_i > \boldsymbol{\widetilde{y}}_i$ and $\boldsymbol{\widetilde{x}}_j = \boldsymbol{\widetilde{y}}_j$ for all $j < i$.
\end{definition}

The offline leximin mechanism chooses an outcome such that the vector of accumulated utilities is maximal according to the leximin ordering. Formally, let $\argmax^{\succ}$ be the maximum elements under the leximin ordering.
\begin{definition}
The offline leximin, denoted by $\textsf{LMin}_{\textsf{off}}$, chooses an outcome such that 
$ \boldsymbol{o} \in \argmax^{\succ}_{\boldsymbol{o'}} ~ u(\boldsymbol{o'})$.
\end{definition}
The leximin mechanisms are not scale-free, and thus a very high (or low) valuation might bias the leximin mechanisms. Therefore, 
we consider two normalization methods, using either the $RRS_i$ or $Prop_i$ values. Let $uRRS(\boldsymbol{o})$ be the vector of utilities $u(\boldsymbol{o})$, in which each $u_i(\boldsymbol{o})$ is divided by $RRS_i$
\footnote{If $RRS_i = 0$, we replace it with  an infinitesimal quantity $\epsilon$ \cite{freeman2017fair}.}.
Similarly, $uProp(\boldsymbol{o})$ is the vector of utilities in which each $u_i(\boldsymbol{o})$ is divided by $Prop_i$.  

\begin{definition}
The offline leximin $RRS$, denoted by $\textsf{LMin}^{\textsf{RRS}}_{\textsf{off}}$, chooses an outcome such that 
$ \boldsymbol{o} \in \argmax^{\succ}_{\boldsymbol{o'}} ~ uRRS(\boldsymbol{o'}) $.
\end{definition}

\begin{definition}
The offline leximin $Prop$, denoted by $\textsf{LMin}^{\textsf{Prop}}_{\textsf{off}}$, chooses an outcome such that 
$ \boldsymbol{o} \in \argmax^{\succ}_{\boldsymbol{o'}} ~ uProp(\boldsymbol{o'})$.
\end{definition}

In the online setting, since the mechanism does not get the entire input upfront, normalizing the input up to a specific round is meaningless. Moreover, since we show that it is impossible to satisfy proportionality or its relaxations with unrestricted valuations (\Cref{theorm:not_prop1,theorm:not_rrs}), we study the online setting only with Borda valuations. 
Therefore, we consider the online leximin mechanism without normalization.
\begin{definition}
The online leximin, denoted by $\textsf{LMin}_{\textsf{on}}$, chooses an outcome $o^t$ for round $t$ such that 
$ o^t \in \argmax^{\succ}_{c_j \in C} ~ (u(\boldsymbol{o}^{t-1}) + \boldsymbol{v^t}(c_j)).$
\end{definition}

Note that the outcome of $\textsf{RR}_{\textsf{off}}$ can be computed efficiently. However, it is intractable to compute the outcomes of $\textsf{MNW}_{\textsf{off}}$ and $\textsf{LMin}^{\textsf{RRS}}_{\textsf{off}}$, due to \cite{conitzer2017fair}. It is also intractable to compute the outcomes of $\textsf{LMin}^{\textsf{Prop}}_{\textsf{off}}$, since $\textsf{LMin}^{\textsf{Prop}}_{\textsf{off}}$ satisfies $MPP$ (\autoref{thm:lex_prop_satisfies_pie}), and it can thus decide whether a proportional outcome exists, which is a computationally hard problem \cite{bouveret2016characterizing}. Clearly, the outcome of all of the online mechanisms can be computed efficiently.

\section{Connections between Fairness Properties}
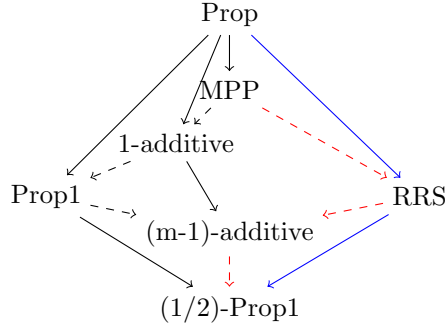
\begin{figure}[htb]
  \centering

\begin{tikzpicture}[node distance=1cm]
    \node (Prop) {Prop};
    \node (MPP) [below of=Prop] {MPP};
    \node (oneAdd) [below left of=MPP] {1-additive};
    \node (RRS) [below right of=oneAdd,xshift=2.5cm] {RRS};
    \node (Prop1) [below left of=oneAdd,xshift=-1cm] {Prop1};
    \node (mMinusOneAdd) [below right of=oneAdd,yshift=-0.5cm] {(m-1)-additive};
    \node (halfProp1) [below of=mMinusOneAdd] {(1/2)-Prop1};

   \draw[->] (Prop) -- (MPP);
   \draw[->] (Prop) -- (Prop1);
   \draw[->,blue] (Prop) -- (RRS);
   \draw[->] (Prop) -- (oneAdd);
   \draw[->] (Prop1) -- (halfProp1);
   \draw[->,blue] (RRS) -- (halfProp1);
   \draw[->,dashed] (MPP) -- (oneAdd);
   \draw[->,dashed,red] (MPP) -- (RRS);
   \draw[->,dashed] (Prop1) -- (mMinusOneAdd);
   \draw[->,dashed,red] (RRS) -- (mMinusOneAdd);
   \draw[->,dashed] (oneAdd) -- (Prop1);
   \draw[->] (oneAdd) -- (mMinusOneAdd);
   \draw[->,dashed,red] (mMinusOneAdd) -- (halfProp1);
\end{tikzpicture}
\caption{A graph depicting the connections between the fairness properties. $x$-additive means $x$-additive approximation of $Prop$. The dashed edges are only valid with Borda valuations. The blue edges are due to \protect\cite{conitzer2017fair}.  The red edges are shown in the appendix.}
\label{Fig:axioms_connection}
\end{figure}

Before analyzing the mechanisms, we show the connections between the various fairness properties of a given outcome, as can be shown in \autoref{Fig:axioms_connection}.

Specifically, \citeauthor{conitzer2017fair}~\cite{conitzer2017fair} show that $Prop$ entails $RRS$, and $RRS$ entails $1/2$-$Prop1$. Clearly, an outcome that satisfies a fairness property satisfies all of its relaxations and approximations. We show additional connections when the valuations are Borda valuations, and we begin with $MPP$. Note that with Borda valuations,  $RRS_i = \floor{\frac{T}{n}}(m-1)$ and $Prop_i = \frac{T(m-1)}{n}$, for every agent $i$.

\begin{theoremrep}
\label{theorem:mpp_entails_1_add}
With Borda valuations, $MPP$ entails $1$-additive approximation of $Prop$.
\end{theoremrep}
\begin{proof}
Suppose that $\boldsymbol{o}$ satisfies $MPP$. Clearly, if $Prop_i \leq 1$, then $\boldsymbol{o}$ satisfies a $1$-additive approximation of $Prop$. If $Prop_i > 1$, let $\boldsymbol{o'}$ be an outcome that satisfies a $1$-additive approximation of $Prop$. Note that such an outcome always exists due to \autoref{theorem:1-additive-prop-leximin}. By definition, $u_i(\boldsymbol{o'}) + 1 \geq Prop_i$ for any agent $i$. That is, $u_i(\boldsymbol{o'}) \geq \frac{Prop_i - 1}{Prop_i} \cdot Prop_i$. Since $Prop_i > 1$, $\frac{Prop_i - 1}{Prop_i} > 0$. In addition, $Prop_i = \frac{T(m-1)}{n}$ for any agent $i$, and we can thus write that $\boldsymbol{o'}$ satisfies $\frac{Prop_i - 1}{Prop_i}$-$Prop$. Now, since $\boldsymbol{o}$ satisfies $MPP$, it must also satisfy $\frac{Prop_i - 1}{Prop_i}$-$Prop$. Therefore, $u_i(\boldsymbol{o}) \geq \frac{Prop_i - 1}{Prop_i} \cdot Prop_i$, which implies that $u_i(\boldsymbol{o}) + 1 \geq Prop_i$. That is, $\boldsymbol{o}$ satisfies a $1$-additive approximation of $Prop$.
\end{proof}

\begin{toappendix}
    
\begin{theorem}
With Borda valuations, $MPP$ entails $RRS$.   
\end{theorem}
\begin{proof}
Suppose that $\boldsymbol{o}$ satisfies $MPP$. Clearly, if $T < n$, then $RRS_i = 0$ for any agent $i$, thus $\boldsymbol{o}$ satisfies $RRS$. If $T \geq n$, let $\boldsymbol{o'}$ be an outcome that satisfies $RRS$. Note that such an outcome always exists due to \cite{conitzer2017fair}. By definition, $u_i(\boldsymbol{o'}) \geq RRS_i = \floor{\frac{T}{n}} (m-1)$ for any agent $i$. That is, $u_i(\boldsymbol{o'}) \geq \floor{\frac{T}{n}} \frac{(m-1)}{Prop_i} Prop_i$. Since  $T \geq n$, $\floor{\frac{T}{n}} > 0$. That is, $\floor{\frac{T}{n}} \frac{(m-1)}{Prop_i} > 0$. In addition, $Prop_i = \frac{T(m-1)}{n}$ for any agent $i$, and we can thus write that $\boldsymbol{o'}$ satisfies $\floor{\frac{T}{n}} \frac{(m-1)}{Prop_i}$-$Prop$. Since $\boldsymbol{o}$ satisfies $MPP$, it must also satisfy $\floor{\frac{T}{n}} \frac{(m-1)}{Prop_i}$-$Prop$. Therefore, $u_i(\boldsymbol{o}) \geq \floor{\frac{T}{n}} \frac{(m-1)}{Prop_i} Prop_i$, which implies that $u_i(\boldsymbol{o}) \geq \floor{\frac{T}{n}} {(m-1)} = RRS_i$. That is, $\boldsymbol{o}$ satisfies $RRS$.
\end{proof}

Next, we consider $RRS$.

\begin{theorem}
With Borda valuations, $RRS$ entails $(m-1)$-additive approximation of $Prop$.   
\end{theorem}
\begin{proof}
Suppose that $\boldsymbol{o}$ satisfies $RRS$. That is, $u_i(\boldsymbol{o}) \geq \floor{\frac{T}{n}} (m-1)$ for any agent $i$. We can add $(m-1)$ to both sides of the inequality to get $u_i(\boldsymbol{o}) + (m-1) \geq \floor{\frac{T}{n}} (m-1) + (m-1)$. It is clear that $\floor{\frac{T}{n}} (m-1) + (m-1) \geq \frac{T}{n}(m-1) = Prop_i$. That is, $u_i(\boldsymbol{o}) + (m-1) \geq Prop_i$, which means that $\boldsymbol{o}$ satisfies $(m-1)$-additive approximation of $Prop$.
\end{proof}
\end{toappendix}

\begin{theoremrep}
\label{theorem:1_add_entails_prop1}
With Borda valuations, $1$-additive approximation of $Prop$ entails $Prop1$.
\end{theoremrep}
\begin{proof}
Suppose that $\boldsymbol{o}$ satisfies a $1$-additive approximation of $Prop$. Clearly, if $\boldsymbol{o}$ satisfies $Prop$, it also satisfies $Prop1$. If $\boldsymbol{o}$ does not satisfy $Prop$, then for any agent $i$ in which $u_i(\boldsymbol{o}) < Prop_i$ there exist a round $t$ such that $v^t_i(o^t) < v^t_i(cMax^t_i)$. 
Since we use Borda valuations, $v^t_i({cMax}^t_i) - v^t_i(o^t) \geq 1$. In addition, since $\boldsymbol{o}$ satisfies a $1$-additive approximation of $Prop$, then $u_i(\boldsymbol{o})  + 1 \geq Prop_i$. Combining the two inequalities, we get that
$(u_i(\boldsymbol{o})  + 1) + (v^t_i({cMax}^t_i) - v^t_i(o^t)) \geq Prop_i + 1$. That is, $\boldsymbol{o}$ satisfies $Prop1$.
\end{proof}

\begin{theoremrep}
\label{theorem:prop1_entails_m-1}
With Borda valuations, $Prop1$ entails $(m-1)$-additive approximation of $Prop$.  
\end{theoremrep}
\begin{proof}
Suppose that $\boldsymbol{o}$ satisfies $Prop1$. That is,  for each agent $i \in N$, there exists a round $t$ such that
$u_i(\boldsymbol{o}) - v^t_i(o^t) + v^t_i({cMax}^t_i) \geq Prop_i.$
With Borda valuations, for any agent $i$ and round $t$, $v^t_i({cMax}^t_i) = m-1$ and $v^t_i(o^t) \geq 0$. Therefore, 
$u_i(\boldsymbol{o}) + (m-1)\geq Prop_i$. That is, $\boldsymbol{o}$ satisfies $(m-1)$-additive approximation of $Prop$.
\end{proof}

\begin{toappendix}
\begin{theorem}
With Borda valuations, $(m-1)$-additive approximation of $Prop$ entails $1/2$-$Prop1$.
\end{theorem}
\begin{proof}
\begin{proof}
Suppose that $\boldsymbol{o}$ satisfies $(m-1)$-additive approximation of $Prop$. If $T \leq 2n$, then $\frac{1}{2}\frac{T}{n} \leq 1$. That is, $(m-1) \geq \frac{1}{2} \frac{T(m-1)}{n} = \frac{1}{2}Prop_i$. In addition, with Borda valuations, by definition, $v^t_i({cMax}^t_i) = (m-1)$, and $v^t_i(o^t) \geq 0$ for any agent $i$ and round $t$. That is, $- v^t_i(o^t) + v^t_i({cMax}^t_i) \geq (m-1)$ for any agent $i$ and round $t$. We thus get that, $u_i(\boldsymbol{o}) - v^t_i(o^t) + v^t_i({cMax}^t_i) \geq \frac{1}{2}Prop_i$ for any agent $i$ and round $t$. That is, $\boldsymbol{o}$ satisfies $1/2$-$Prop1$. If $T > 2n$, then, by definition, $u_i(\boldsymbol{o}) + (m-1) \geq Prop_i$ for any agent $i$. In addition, $Prop_i = \frac{T(m-1)}{n}$ for any agent $i$. That is, $u_i(\boldsymbol{o}) + (m-1) \geq \frac{T(m-1)}{n}$. Therefore, $u_i(\boldsymbol{o}) \geq \frac{(T - n)(m-1)}{n}$. In addition, since $T \geq 2n$, then $2(T-n) \geq T$. We thus get that, $u_i(\boldsymbol{o}) \geq \frac{(T - n)(m-1)}{n} = \frac{2(T - n)(m-1)}{2n} \geq \frac{T(m-1)}{2n} = \frac{1}{2}Prop_i$. Therefore, $\boldsymbol{o}$ satisfies $1/2$-$Prop$, and thus satisfies $1/2$-$Prop1$.
\end{proof}
\end{proof}
\end{toappendix}

\section{Offline Setting}
We begin by showing that $\textsf{RR}_{\textsf{off}}$, $\textsf{MNW}_{\textsf{off}}$, and $\textsf{LMin}^{\textsf{RRS}}_{\textsf{off}}$ might not find a proportional outcome, even if such an outcome exists. Specifically, consider the following examples.
\begin{example}
\label{exmp:rr_dont_find_prop}
Let $m = 6, n = 2, T = 1$, and  
$V^1 = \begin{pmatrix}
5 & 4 & 3 & 2 & 1 & 0\\
2 & 1 & 3 & 0 & 4 & 5
\end{pmatrix}
.$
The only outcome that satisfies $Prop$ is $\boldsymbol{o} = (c_3)$. However, $\textsf{MNW}_{\textsf{off}}$ chooses $(c_1)$, and $\textsf{RR}_{\textsf{off}}$ chooses either $(c_1)$ or $(c_6)$.
\end{example}
\begin{example}
Let $m = 3, n = 2,T = 2$, and  
$V^1 = \begin{pmatrix}
5000 & 2500 & 50\\
30 & 40 & 50
\end{pmatrix}, \quad
V^2 = \begin{pmatrix}
0 & 1 & 0\\
0 & 1 & 0
\end{pmatrix}.$
The outcome $\boldsymbol{o} = (c_2,c_2)$ satisfies $Prop$. However, since $RRS_1 = $RRS$_2 = 1$, then   $\textsf{LMin}^{\textsf{RRS}}_{\textsf{off}}$ chooses the candidate $c_3$ in the first round.    
\end{example}
It is not a coincidence that $\textsf{RR}_{\textsf{off}}$, $\textsf{MNW}_{\textsf{off}}$, and $\textsf{LMin}^{\textsf{RRS}}_{\textsf{off}}$ might fail to find a proportional outcome. We claim that, in general, these mechanisms might output an outcome that is ``far'' from being proportional, 
as formally captured by the $MPP$ property. 
Moreover, note that $\textsf{RR}_{\textsf{off}}$ and $\textsf{LMin}^{\textsf{RRS}}_{\textsf{off}}$ satisfy $RRS$, and $\textsf{MNW}_{\textsf{off}}$ satisfies $Prop1$ \cite{conitzer2017fair}. We show unavoidable tradeoffs between $MPP$ and $\alpha$-$RRS$, for any constant $\alpha$, and between $MPP$ and $\alpha$-$Prop1$, for any constant $\alpha$.

The proof is based on the intuition that both $Prop1$ and $RRS$ might fail to find a proportional outcome when there is a round $t$ in which an agent $i$ assigns a relatively high valuation to one of the candidates. In this case, $Prop1$ can be satisfied while ignoring the valuations (of agent $i$) in the other rounds, which might result in an unproportional outcome. $RRS_i$, in this case, does not depend on the valuation of $i$ in round $t$ (since the rounds are sorted in a non-ascending order), which might also result in an unproportional outcome. However, the $MPP$ fairness property must consider the valuations of all the agents, in all of the rounds.   

\begin{theoremrep} \label{thm:lex_prop_off_not_approx}
If $n > 2$, a mechanism that satisfies $MPP$ does not satisfy $\alpha$-$RRS$, for any constant $\alpha$, and it does not satisfy $\alpha$-$Prop1$, for any constant $\alpha$. 
\end{theoremrep}
\begin{proofsketch}
We build a scenario with $n$ agents, and three distinct agents, $i$ ,$j$, and $k$. In the first round, agents $j$ and $k$ assign very high valuations to different candidates, while in all other rounds they assign very low valuations to all the candidates. Note that it is impossible to satisfy both $j$ and $k$ in the first round, and assume that the candidate favored by $j$ is selected in the first round. A mechanism that satisfies $MPP$ must compensate $k$ in the other rounds. As a result, the utility of agent $i$ is reduced, but still in a way that guarantees her some minimal utility, and overall, all three agents receive this minimal utility. 
In contrast, a mechanism that satisfies $Prop1$ or $RRS$ ignores the first round due to the nature of these fairness properties and, as a result, agent $i$ ends up with a utility that is not very low, while agent $k$ ends up with a much lower utility.
\end{proofsketch}
\begin{proof}
Let $\epsilon \in R$ such that $\frac{1}{T(T-1)} > \epsilon > 0$. That is, 
$ -\epsilon + \frac{1}{(T - 1)} > \epsilon(T - 1)$.
Let $m,n > 2$, $T \geq 2n$, and  
$$v_1^t(c_j) = \begin{cases}
    \epsilon(T - 1),& \text{if } t = T \land  j=1\\
    -\epsilon + \frac{1}{(T - 1)},& \text{if } t \not= T \land  j=1\\
    0,              & \text{otherwise}
\end{cases}$$
$$ v_2^t(c_j) = \begin{cases}
    1,&\text{if } t = 1\land j=2\\
    0,              & \text{otherwise}
\end{cases}  \quad v_{i \not\in\{1,2,3\}}^t(c_j) = \frac{1}{T} $$

$$v_3^t(c_j) = \begin{cases}
    1 - \epsilon(T-1),&\text{if }  t = 1 \land j=3\\
    2\epsilon,&\text{if }  t = 2 \land j=3\\
    \epsilon,&\text{if }  t \not\in \set{1,2,T} \land j=3\\
    0,              & \text{otherwise}
\end{cases}.$$

That is, the valuations in matrix form are as follows:
$$V^1 = \begin{pmatrix}
-\epsilon + \frac{1}{(T-1)} & 0 & 0 & 0 & \ldots\\
0 & 1 & 0 & 0 & \ldots\\
0 & 0 & 1-\epsilon(T-1) & 0 & \ldots\\
\frac{1}{T} & \frac{1}{T} & \frac{1}{T} & \frac{1}{T} & \ldots\\
\vdots & \vdots & \vdots & \ddots 
\end{pmatrix}$$ $$ V^2 = \begin{pmatrix}
-\epsilon + \frac{1}{(T-1)} & 0 & 0 & 0 & \ldots\\
0 & 0 & 0 & 0 & \ldots\\
0 & 0 & 2\epsilon & 0 &  \ldots\\
\frac{1}{T} & \frac{1}{T} & \frac{1}{T} & \frac{1}{T} & \ldots\\
\vdots & \vdots & \vdots  &\ddots 
\end{pmatrix}
$$

$$V^t = \begin{pmatrix}
-\epsilon + \frac{1}{(T-1)} & 0 & 0 & 0 & \ldots\\
0 & 0 & 0 & 0 & \ldots\\
0 & 0 & \epsilon & 0 & \ldots\\
\frac{1}{T} & \frac{1}{T} & \frac{1}{T} & \frac{1}{T} & \ldots\\
\vdots & \vdots & \vdots  &\ddots 
\end{pmatrix} $$ $$ V^T = \begin{pmatrix}
\epsilon{(T-1)} & 0 & 0 & 0 & \ldots\\
0 & 0 & 0 & 0 & \ldots\\
0 & 0 & 0 & 0 & \ldots\\
\frac{1}{T} & \frac{1}{T} & \frac{1}{T} & \frac{1}{T} & \ldots\\
\vdots & \vdots & \vdots  &\ddots 
\end{pmatrix}
.$$

We first show that for any agent $i$, $Prop_i = \frac{1}{n}$.
By definition, for any agent $i$, $Prop_i = \frac{1}{n}\sum_{t=1}^T v^t_i({cMax}^t_i)$. That is, 
$Prop_1 = \frac{1}{n}(\epsilon(T - 1) + (-\epsilon + \frac{1}{(T - 1)})(T-1)) = \frac{1}{n}$, 
$Prop_2 = \frac{1}{n}$, and
$Prop_3 = \frac{1}{n}((1- \epsilon(T-1) + 2\epsilon + \epsilon(T-3)) = \frac{1}{n}$.
For any other agent $i \not\in \set{1,2,3}$, $Prop_i = \frac{1}{n} (T\frac{1}{T}) =  \frac{1}{n}$.

Let $\boldsymbol{o}$ be the outcome that is chosen by a mechanism that satisfies $MPP$. 
Let $$\hat{\boldsymbol{o}} =  \begin{cases}
    c_1,&\text{if }  t = T\\
    c_2,&\text{if }  t = 1\\
    c_3,              & \text{otherwise}
\end{cases}.$$
That is, 
$$u_i(\hat{\boldsymbol{{o}}}) =\begin{cases}
    \epsilon(T-1),&\text{if }  i \in \{1,3\}\\
    1,              & \text{otherwise}
\end{cases}.$$
Note that $1 > \frac{1}{2} > \frac{n(T-1)}{T(T-1)} > n\epsilon(T-1) > \epsilon(T-1) > 0$, and recall that for any agent $i$, $Prop_i = \frac{1}{n}$.
Therefore, for any agent $i$, $u_{i}(\hat{\boldsymbol{o}}) \geq \epsilon(T-1) = \frac{\epsilon(T-1)n}{n} = (n\epsilon(T-1))Prop_i$.
That is $\hat{\boldsymbol{o}}$ satisfies $n\epsilon(T-1)$-Prop. Since $\boldsymbol{o}$ satisfies $MPP$, $\boldsymbol{o}$ also satisfies $n\epsilon(T-1)$-Prop. That is, for any agent $i$, $u_i(\boldsymbol{o}) \geq (n\epsilon(T-1))Prop_i > 0$. 

Let's examine the choice of the candidates in $\boldsymbol{o}$. 
Since $u_2(\boldsymbol{o}) > 0$, and agent $2$ assigns a positive valuation only for candidate $c_2$ and only in round $1$, then ${o}^1 = c_2$.
Now, since $u_3(\boldsymbol{o}) \geq \epsilon(T-1)$ and we showed that ${o}^1 = c_2$, it must be that ${o}^t = c_3$ for every round $t$,  $1 < t < T$.
Finally, since $u_1(\boldsymbol{o}) \geq \epsilon(T-1)$, it must be that ${o}^T = c_1$. Overall, $\boldsymbol{o} = \hat{\boldsymbol{o}}$.

We now show that a mechanism that satisfies $MPP$ does not satisfy $\alpha$-$RRS$, for any constant $\alpha < 1$.
Assume by contradiction that there is a constant $0 < \alpha < 1$ such that $\boldsymbol{o}$ satisfy $\alpha$-$RRS$. Therefore, $\epsilon(T-1) = u_1(\boldsymbol{o}) \geq \alpha \cdot RRS_1$.
Since $T \geq 2n$, then $RRS_1 \geq cMax_1(n) = (-\epsilon + \frac{1}{(T - 1)})$.
Therefore, $\epsilon(T-1) \geq \alpha \cdot (-\epsilon + \frac{1}{(T - 1)})$.
Let $\epsilon = \frac{\alpha}{T(T-1)}$, and note that since $0 < \alpha < 1$ then $0 < \epsilon < \frac{1}{T(T-1)}$. 
We thus get that $\frac{\alpha}{T(T-1)}(T-1) \geq \alpha \cdot (-\frac{\alpha}{T(T-1)} + \frac{1}{(T - 1)})$. That is, 
 $(T-1) \geq (-\alpha + T)$, and thus $\alpha \geq 1$, a contradiction.

Next, we show that a mechanism that satisfies $MPP$ does not satisfy $\alpha$-$Prop1$, for any constant $\alpha < 1$.
Assume by contradiction that there is a constant $0 < \alpha < 1$ such that $\boldsymbol{o}$ satisfy $\alpha$-$Prop1$.
Therefore, there exists $t$ such that  $u_1(\boldsymbol{o}) - v^t_1(o^t) + v^t_1({cMax}^t_1) \geq \alpha \cdot Prop_1$. 
Since $Prop_i = \frac{1}{n}$, $u_1(\boldsymbol{o}) - v^t_1(o^t) + v^t_1({cMax}^t_1) \geq \alpha \cdot \frac{1}{n}$. Moreover, $u_1(\boldsymbol{o}) - v^t_1(o^t) + v^t_1({cMax}^t_1) \leq u_1(\boldsymbol{o}) + v^t_1({cMax}^t_1) \leq \epsilon(T-1) - \epsilon + \frac{1}{(T - 1)}$, since $ -\epsilon + \frac{1}{(T - 1)} > \epsilon(T - 1)$.
That is, $\epsilon(T-2) + \frac{1}{(T - 1)} \geq \alpha \cdot \frac{1}{n}$.
Let $\epsilon = \alpha (\frac{\alpha(T-1) - n}{nT(T-1)(T-2)})$, and  $T \geq \frac{2n}{\alpha} + 1$.
Note that since $0 < \alpha < 1$, and  $T \geq \frac{2n}{\alpha} + 1 \geq 2n$,
then $T \geq \alpha(T-1) - n \geq \alpha(\frac{2n}{\alpha} + 1-1) - n = n > 0$. 
We thus get that $(\frac{1}{T(T-1)}) \geq (\frac{1}{3(T-1)(T-2)}) > \alpha (\frac{1}{n(T-1)(T-2)}) > \alpha \frac{\alpha(T-1) - n}{T} (\frac{1}{n(T-1)(T-2)}) = \epsilon$. In addition, $\epsilon = \alpha (\frac{\alpha(T-1) - n}{nT(T-1)(T-2)}) \geq \alpha(\frac{n}{nT(T-1)(T-2)}) > 0$.
Overall, $\frac{1}{T(T-1)} > \epsilon > 0$. 
We thus get that $\alpha (\frac{\alpha(T-1) - n}{nT(T-1)(T-2)})(T-2) + \frac{1}{(T - 1)}  \geq \alpha \cdot \frac{1}{n}$. That is, $\alpha {(\alpha(T-1) - n)} + nT  \geq \alpha \cdot T(T-1)$, and thus $n(T - \alpha)  \geq \alpha (T-1) (T - \alpha) $. That is, $n \geq \alpha(T-1) \geq \alpha(\frac{2n}{\alpha} + 1 - 1) = 2n$, a contradiction.

Finally, note that a mechanism that satisfies $RRS$ also satisfies $\alpha$-$RRS$ for any $\alpha < 1$, and a mechanism that satisfies $Prop1$ also satisfies $\alpha$-$Prop1$ for any $\alpha < 1$. Therefore, a mechanism that satisfies $MPP$ does not satisfy $\alpha$-$RRS$, for any constant $\alpha$, and it does not satisfy $\alpha$-$Prop1$, for any constant $\alpha$. 
\end{proof}

Since we show that $\textsf{RR}_{\textsf{off}}$, $\textsf{MNW}_{\textsf{off}}$, and $\textsf{LMin}^{\textsf{RRS}}_{\textsf{off}}$ do not satisfy $MPP$ even with $2$ agents, we introduce $\textsf{LMin}^{\textsf{Prop}}_{\textsf{off}}$, which satisfies $MPP$.

\begin{theoremrep}
\label{thm:lex_prop_satisfies_pie}
$\textsf{LMin}^{\textsf{Prop}}_{\textsf{off}}$ satisfies $MPP$.
\end{theoremrep}
\begin{proof} 
Suppose that there exists an $\alpha$ with which there is an outcome $\boldsymbol{o'}$ that satisfies $\alpha$-$Prop$. That is, the minimal utility value in $uProp(o')$ is at least $\alpha$.
By definition, the outcome $\boldsymbol{o}$ that is returned by $\textsf{LMin}^{\textsf{Prop}}_{\textsf{off}}$ satisfies $uProp(o) \succeq uProp(o')$.  Therefore, the minimal utility value in $uProp(o)$ is at least $\alpha$. That is, $o$ satisfies $\alpha$-$Prop$. Therefore, $\textsf{LMin}^{\textsf{Prop}}_{\textsf{off}}$ satisfies $MPP$.
\end{proof}

Clearly, $\textsf{LMin}^{\textsf{Prop}}_{\textsf{off}}$ also satisfies Pareto optimality. However, since $\textsf{LMin}^{\textsf{Prop}}_{\textsf{off}}$ satisfies $MPP$, it does not satisfy $\alpha$-$RRS$ or $\alpha$-$Prop1$, for any constant $\alpha$. Indeed, the proof of ~\autoref{thm:lex_prop_off_not_approx} assumes that the number of agents, $n$, is at least $3$. If there are only two agents, $\textsf{LMin}^{\textsf{Prop}}_{\textsf{off}}$ satisfies $RRS$ and $1/2$-$Prop1$.

\begin{theoremrep}
$\textsf{LMin}^{\textsf{Prop}}_{\textsf{off}}$ with 2 agents satisfies $RRS$ and  $1/2$-$Prop1$.
\end{theoremrep}
\begin{proof}
Clearly, in $\textsf{RR}_{\textsf{off}}$, the first agent according to the turn's order, $\pi$, receives a utility that is at least his $Prop$ value. That is, if agent $1$ is the first agent according to $\pi$, then $\textsf{RR}_{\textsf{off}}$ returns an outcome that guarantees a utility of at least $Prop_1$ to agent $1$, and a utility of at least $RRS_2$ to agent $2$. Similarly, if agent $2$ is the first agent according to $\pi$, then $\textsf{RR}_{\textsf{off}}$ returns an outcome that guarantees a utility of at least $Prop_2$ to agent $2$, and a utility of at least $RRS_1$ to agent $1$. Therefore, the outcome $\boldsymbol{o}$ that is returned by $\textsf{LMin}^{\textsf{Prop}}_{\textsf{off}}$ satisfies $uProp(\boldsymbol{o}) \succeq (1,\frac{RRS_2}{Prop_2})$ and $ uProp(\boldsymbol{o}) \succeq (\frac{RRS_1}{Prop_1},1)$. Now, since $Prop_i \geq RRS_i$ \cite{conitzer2017fair}, and there are exactly two agents, then $uProp(\boldsymbol{o}) \succeq (\max(\frac{RRS_2}{Prop_2},\frac{RRS_1}{Prop_1}),\max(\frac{RRS_2}{Prop_2},\frac{RRS_1}{Prop_1}))$ 
Hence, $uProp_i(\boldsymbol{o}) \geq \max(\frac{RRS_2}{Prop_2},\frac{RRS_1}{Prop_1}) \geq \frac{RRS_i}{Prop_i}$ and thus $u_i(\boldsymbol{o}) \geq RRS_i$. That is, $\textsf{LMin}^{\textsf{Prop}}_{\textsf{off}}$ with two agents satisfies $RRS$.
Finally, due to \cite{conitzer2017fair}, $RRS$ entails $1/2$-$Prop1$.
\end{proof}
That is, when there are only two agents, $\textsf{LMin}^{\textsf{Prop}}_{\textsf{off}}$ is (arguably) the ``fairest''  mechanism among the mechanisms that we study, since it is the only one that satisfies $PO$, $1/2$-$Prop1$, $MPP$, and $RRS$, simultaneously. 

\subsection{Borda Valuations}
Recall that with Borda valuations, $RRS_i$ is the same for all $i$. Similarly, $Prop_i$ is the same for all $i$. Thus, in this setting, $\textsf{LMin}^{\textsf{Prop}}_{\textsf{off}}$ and $\textsf{LMin}^{\textsf{RRS}}_{\textsf{off}}$ are equivalent to $\textsf{LMin}_{\textsf{off}}$. That is, there is effectively a single leximin mechanism, $\textsf{LMin}_{\textsf{off}}$, with Borda valuations. 

Clearly, every mechanism that satisfies an efficiency or fairness property without any restriction on the valuation, satisfies the property with Borda valuations. We thus analyze the properties that are not satisfied in the general case, and we begin with $Prop$.

Since $\textsf{LMin}_{\textsf{off}}$ satisfies $MPP$ (\autoref{thm:lex_prop_satisfies_pie}), then with Borda valuations, $\textsf{LMin}_{\textsf{off}}$ guarantees a $1$-additive approximation of $Prop$ (\autoref{theorem:mpp_entails_1_add}). 
Moreover, we show that $\textsf{LMin}_{\textsf{off}}$ guarantees the best possible constant-factor approximation of $Prop$.
\begin{theoremrep}
Even with Borda valuations, there is no mechanism that guarantees an additive constant-factor approximation of $Prop$ better than $1$.
\end{theoremrep}
\begin{proof}
Assume by contradiction that there is a mechanism that guarantees an $\alpha$-additive approximation of $Prop$, with $\alpha < 1$. Since $\alpha < 1$, we can choose $n$ such that $\frac{n-1}{n} > \alpha$. Let $T = 1$ and $m = n$. In addition, assume that the agents use Borda valuations, and let $v_i^1(c_i) = 0$, for any agent $i$. Given any outcome $\boldsymbol{o}$, there must be an agent $j$ that receives a utility of $0$ from $\boldsymbol{o}$. On the hand, $Prop_j = \frac{n-1}{n}$. That is, the difference between $Prop_j$ and $u_j(\boldsymbol{o})$ is strictly greater than $\alpha$. A contradiction.
\end{proof}

We now show that both $\textsf{RR}_{\textsf{off}}$ and $\textsf{MNW}_{\textsf{off}}$ do not guarantee an additive constant-factor approximation of $Prop$. Specifically, $\textsf{RR}_{\textsf{off}}$ does not guarantee an additive approximation that is better than $m-1$. 
\begin{theoremrep}
\label{theorem:rr_not_m-1_app}
With Borda valuations, $\textsf{RR}_{\textsf{off}}$ does not guarantee an additive approximation of $Prop$ that is better than $m-1$. 
\end{theoremrep}
\begin{proof}
Assume by contradiction that $\textsf{RR}_{\textsf{off}}$ guarantees an $(m-1 - \alpha)$-additive approximation of $Prop$, with $\alpha >0$. Since $\alpha >0$, we can choose $n$ such that $\frac{(n-1)(m-1)}{n} > m-1 - \alpha$.
Let $T = n-1$ and for every round $t$, and let $$V^t =  \begin{pmatrix}
m-1 & m-2 & \ldots & 0\\
\vdots \\
m-1 & m-2 & \ldots & 0\\
0 & 1 & \ldots & m-1\\
\end{pmatrix}.$$
Assume that $\textsf{RR}_{\textsf{off}}$ uses the identity permutation as the order over the agents. Clearly, $\boldsymbol{o}=(c_1, \ldots, c_1)$, and thus $u_n(\boldsymbol{o})=0$. On the other hand, $Prop_n = \frac{(n-1)(m-1)}{n}$. That is, the difference between $Prop_n$ and  $u_n(\boldsymbol{o})$ is strictly greater than $m-1 - \alpha$. A contradiction.
\end{proof} 

Indeed, $\textsf{RR}_{\textsf{off}}$ satisfies $Prop1$ \cite{conitzer2017fair}, and this entails that $\textsf{RR}_{\textsf{off}}$ guarantees an $(m-1)$-additive approximation of $Prop$, as we show in \autoref{theorem:1_add_entails_prop1}.
Similarly, since $\textsf{MNW}_{\textsf{off}}$ satisfies $Prop1$, $\textsf{MNW}_{\textsf{off}}$ also guarantees an $(m-1)$-additive approximation of $Prop$. On the other hand, $\textsf{MNW}_{\textsf{off}}$ does not guarantee an additive approximation that is better than $(m-3)/2$.

\begin{theoremrep}
\label{theorem:mnw_not_m-5_app}
    Even with Borda valuations, $\textsf{MNW}_{\textsf{off}}$ does not guarantee an additive approximation of $Prop$ that is better than $\frac{m-3}{2}$.  
\end{theoremrep}

\begin{proof}

To prove the theorem, we need the following lemma, which helps to determine the outcome chosen by $MNW$.

\begin{lemma}
\label{lemma:mnw_not_m-5_app}
For any $k,T,n \in \mathbb{N}$, let $f_k$ be a function with a domain $[k(n+1),(k+1)(n+1)] \in \mathbb{R}$, such that $f_k(x) = (k + T(n+1) - x){x}^{n-1}$. let $f$ be a picewise function with a domain $[0,T(n+1)] \in \mathbb{N}$, such that $f(x) = f_k(x)$ if $x \in [k(n+1),(k+1)(n+1))$ (see Figure \ref{Fig:func} for a plot of $f$).
Let $f^{max} = \argmax_{x} f(x)$.
If $n > 2$ and $0 < T \leq \frac{n(n-1)}{2}$ then $f^{max} = \set{T(n+1)}$.
\end{lemma}
\begin{figure}[!htb]

  \centering
 \begin{tikzpicture}
\begin{scope}[xshift=8cm,scale=0.3]
        \draw[->] (0,0) -- (16.2,0) node[right] {$x$};
        \draw[->] (0,0) -- (0,16.2) node[above] {$y$};
        \foreach \x in {0,2,...,16} {
        \draw (\x,0.2) -- (\x,-0.2) node[below] {\x};
        \draw (0.2,\x) -- (-0.2,\x) node[left] {\pgfmathparse{int(\x*30)}\pgfmathresult};
        }

        \clip (0,0) rectangle (16,16);

        \draw[gray, thin] (0,0) grid (16,16);
        \draw[black, thick] (0,0) rectangle (16,16);

        \foreach \a in {0,1,2} {
            \draw[line width=2pt, blue, 
              postaction={decorate, decoration={text along path, text align=center, raise=1ex, text color=blue, text={|\bf|$k = \a$}}}] plot[domain=\a*4:\a*4 + 4] (\x,{(1/30) * (((\a + 12 - \x)*(\x)^2))});

            \ifnum\a>0
                \filldraw[blue] (\a*4,{(1/30) * (((\a + 12 - \a*4)*(\a*4)^2))}) circle (5.5pt) ;
            \fi
            
            \filldraw[blue, draw=blue, fill=white,line width=1.75pt] (\a*4 + 4,{(1/30) * (((\a + 12 - \a*4 - 4)*(\a*4 + 4)^2))}) circle (5pt);
        }

        \foreach \a in {3} {
        \draw[line width=2pt, red, dashed,
              postaction={decorate, decoration={text along path, text align=center, raise=1ex, text color=red, text={|\bf|$k = \a$}}}] plot[domain=\a*4:\a*4 + 3.2] (\x,{(1/30) * (((\a + 12 - \x)*(\x)^2))});
            \filldraw[blue] (\a*4,{(1/30) * (((\a + 12 - \a*4)*(\a*4)^2))}) circle (5.5pt) ;
         }

        
        
    \end{scope}
\end{tikzpicture}
\caption{A plot of the function $f$ when $n = 3$ and $T = \frac{n(n-1)}{2}$. The red dashed line shows the function $f_T$, but since the domain of $f$ is $[0,12]$, $f_T(12)$ is the only point in which $f = f_T$.}
 \label{Fig:func}
\end{figure}
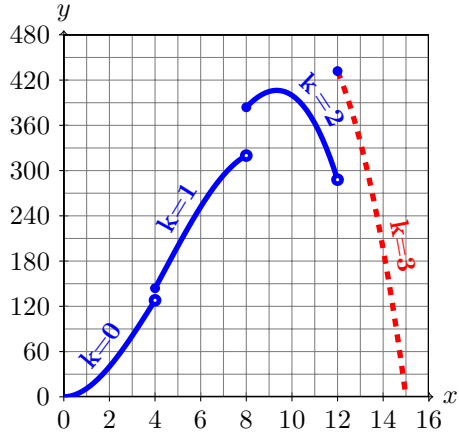
\begin{nestedproof}
Let $f_k^{max} = \argmax_{x \in [k(n+1),(k+1)(n+1)]} f_k(x)$, and let $S_k$ be the set of corresponding stationary points, i.e.,  $S_k = \set{x \mid f'(x) = 0}$.
Assume that $T \leq \frac{n(n-1)}{2}$ and $m = n+2$.
Since for any $k \in \mathbb{N}$, $f_k$ is a  differentiable function, and the domain of $f_k$ is closed, then $f_k^{max} \subseteq S_k \cup \set{k(n+1),(k+1)(n+1)}$.
Now, $f_k'(x) = {x}^{n-2}((n-1)(k + T(n+1)) - n{x})$, and  $f_k''(x) = (n-1){x}^{n-3}((n-2)(k + T(n+1)) - n{x})$.
If $x \in S_k$, $f_k'(x) = 0$ be definition. That is, $x = \frac{(n-1)(k + T(n+1))}{n}$, and $f''(x) < 0$. Therefore,  
$x$ is a maximum point for $f_k$.
However, $S_k$ may be empty. Indeed, we show that $S_k$ is not empty only when $k = T-1$.

Consider $x \in S_k$.
Since $f_k$ is a function with a domain $[k(n+1),(k+1)(n+1)]$, then $k(n+1) \leq x \leq (k+1)(n+1)$. That is, 
$k(n+1) \leq x$ and thus $k(n+1) \leq \frac{(n-1)(k + T(n+1))}{n}$. Thus, $k \leq \frac{T(n^2-1)}{n^2 + 1}$. Since $k \in \mathbb{N}$ it holds that $k \leq \floor{\frac{T(n^2-1)}{n^2 + 1}}$.
On the other hand, $x \leq (k+1)(n+1)$ and thus $\frac{(n-1)(k + T(n+1))}{n} \leq (k+1)(n+1)$. That is, $k \geq \frac{(n+1)(T(n-1) - n)}{n^2 + 1}$, and since $k \in \mathbb{N}$ it holds that $k \geq \ceil{\frac{(n+1)(T(n-1) - n)}{n^2 + 1}}$. Overall, $\floor{\frac{T(n^2-1)}{n^2 + 1}} \geq \ceil{\frac{(n+1)(T(n-1) - n)}{n^2 + 1}}$.
That is, $\floor{T - \frac{2T}{n^2+1}} \geq \ceil{T - 1 - \frac{2T + (n-1)}{n^2 + 1}}$. Since $ 0 < T \leq \frac{n(n-1)}{2}$, then $\floor{T - \frac{2T}{n^2+1}} = T-1$ and $\ceil{T - 1 - \frac{2T + (n-1)}{n^2 + 1}} = T - 1$. That is, $k = T - 1$.

Since we show that $S_k$ is not empty only when $k = T-1$, then for any other $k$, $f_k^{max}$ is either $k(n+1)$ or $(k+1)(n+1)$. If $k < T-1$ then $f_k'(x) > 0$, and if $k > T-1$ then $f_k'(x) < 0$. That is,  
$$f_k^{max} =  \begin{cases} 
        (k+1)(n+1), &  k < T - 1 \\
        \frac{(n-1)(k + T(n+1))}{n}, & k = T - 1 \\  
      k(n+1), & k > T - 1
   \end{cases}.$$

We now concentrate on the points $f_k^{max}$, where $ 0 \leq k \leq T$.
Let $k=T$, and let $\hat{k}=T-1$. We show that $f_{k}(f_k^{max}) > f_{\hat{k}}(f_{\hat{k}}^{max})$. 
Assume by contradiction that $f_{k}(f_k^{max}) \leq f_{\hat{k}}(f_{\hat{k}}^{max})$.
That is, $(k + T(n+1) - k(n+1)){(k(n+1))}^{n-1} \leq (\hat{k} + T(n+1) - \frac{(n-1)(\hat{k} + T(n+1))}{n}){(\frac{(n-1)(\hat{k} + T(n+1))}{n})}^{n-1}$.
Since $k=T$ and $\hat{k}=T-1$, we get that $ T^n{(n+1)}^{n-1} \leq (\frac{(T(n+2) - 1)}{n}){(\frac{(n-1)(T(n+2) - 1)}{n})}^{n-1}$.
Thus, ${T^n}{n^n}{(n+1)}^{n-1} \leq {(n-1)}^{n-1}{{(T(n+2) - 1)}}^{n}$.
That is, ${n^n}{(n+1)}^{n-1} \leq {(n-1)}^{n-1}{{((n+2) - \frac{1}{T})}}^{n}$.
Since $0 < T \leq \frac{n(n-1)}{2}$
then ${(n-1)}^{n-1}{{((n+2) - \frac{1}{T})}}^{n} \leq {(n-1)}^{n-1}{{((n+2) - \frac{2}{n(n-1)})}}^{n}$. 
We thus get that ${n^n}{(n+1)}^{n-1} \leq {(n-1)}^{n-1}{{((n+2) - \frac{2}{n(n-1)})}}^{n}$. 
That is, ${n^n}{(n+1)}^{n-1} \leq {(n-1)}^{n-1}{{(\frac{(n+1)(n^2-2)}{n(n-1)})}}^{n}$. 
We thus get that $(n-1) \leq (n+1){(1-\frac{2}{n^2})}^{n}$. 
That is,  ${(1+\frac{2}{n^2 - 2})}^{n} \leq  \frac{(n+1)}{(n-1)}$.
Recall that the binomial series $(1+x)^\alpha = \sum^{\alpha}_{i = 0} \binom{\alpha}{i}{x^i}$, for any $x \in R,\alpha \in \mathbb{N}$.
Note that $n > 2$ and thus ${(\frac{2}{n^2 - 2})}^i > 0$, for any $i$. Therefore, 
$\sum^{2}_{i = 0} \binom{n}{i}{(\frac{2}{n^2 - 2})}^i = 1 + n(\frac{2}{n^2 - 2}) + \frac{n(n-1)}{2}{(\frac{2}{n^2 - 2})}^2 \leq {(1+\frac{2}{n^2 - 2})}^{n}$.
That is, $1 + n(\frac{2}{n^2 - 2}) + \frac{n(n-1)}{2}{(\frac{2}{n^2 - 2})}^2 \leq \frac{(n+1)}{(n-1)}$.
On the other hand, $1 + n(\frac{2}{n^2 - 2}) + \frac{n(n-1)}{2}{(\frac{2}{n^2 - 2})}^2 = \frac{(n+1)}{(n-1)} + \frac{6n - 8}{(n-1){(n^2-2)}^2}$, and since $n > 2$,   $\frac{(n+1)}{(n-1)} + \frac{6n - 8}{(n-1){(n^2-2)}^2}  > \frac{(n+1)}{(n-1)}$, a contradiction.
Therefore, $f_{k}({f_{k}}^{max}) > f_{\hat{k}}(f_{\hat{k}}^{max})$.

By definition, for any $k \in \mathbb{N}$, $f_{k}(k(n+1)) = 
(k + T(n+1) - k(n+1)){(k(n+1))}^{n-1}$. Clearly, $(k + T(n+1) - (k)(n+1)){(k(n+1))}^{n-1} > 
(k - 1 + T(n+1) - k(n+1)){(k(n+1))}^{n-1} = f_{k - 1}(k(n+1))$. Therefore, $f_{k}(k(n+1)) > f_{k - 1}(k(n+1))$.
We show that if $0 \leq k \leq T-1$ and $\hat{k}=k-1$, then  $f_k(f_k^{max}) > f_{\hat{k}}(f_{\hat{k}}^{max})$. Indeed, $f_{k}(f_{k}^{max}) > f_{k}(k(n+1)) >  f_{k - 1}(k(n+1)) =  f_{\hat{k}}(f_{\hat{k}}^{max})$. Overall, we get that if $k=T$ and $\hat{k} < k$ then $f_k(f_k^{max}) > f_{\hat{k}}(f_{\hat{k}}^{max})$. Moreover, since $f(f_k^{max}) = f_k(f_k^{max})$ then $f^{max} = f_k^{max} =  T(n+1)$.

\end{nestedproof}

We are now ready to prove the theorem.
Let $n >2, m = n+2$, and $T \leq \frac{n(n-1)}{2}$. For any $t \leq T$, let 
$$
V^t = \begin{pmatrix}
m-1 & m-2 & \ldots & 2 & 0 & 1\\
0 & 1 & \ldots & m-3 & m-2 & m-1\\
\vdots \\
0 & 1 & \ldots & m-3 & m-2 & m-1
\end{pmatrix}.$$
Clearly, for any outcome $\boldsymbol{o}$, for any $i,j>1$, $u_i(\boldsymbol{o}) = u_j(\boldsymbol{o})$. We thus concentrate on $u_1(\boldsymbol{o})$ and $u_2(\boldsymbol{o})$.
Note that for any candidate $c_j$, $v^t_1(c_j) = m - j + \mathbbm{1}_{j = m} - \mathbbm{1}_{j = m-1}$, and $v^t_2(c_j) = j-1$. That is, $v^t_1(c_j) + v^t_2(c_j) = (m-1) + \mathbbm{1}_{j = {m}} - \mathbbm{1}_{j = {m-1}}$.
Now, let $count_{c_m}(\boldsymbol{o}) = |\set{t \mid {o}^t = c_{m}}|$, and
$count_{c_{m-1}}(\boldsymbol{o}) = |\set{t \mid {o}^t = c_{m-1}}|$.
That is, $u_1(\boldsymbol{o}) + u_2(\boldsymbol{o}) = \sum_{t=1}^{T} v^t_1({{o}}^t) + \sum_{t=1}^{T} v^t_2({{o}}^t) = \sum_{t=1}^{T} ((m-1) + \mathbbm{1}_{{{o}}^t = c _{{m}}} - \mathbbm{1}_{{{o}}^t = c_{{m-1}}}) = T(m-1) + count_{c_m}(\boldsymbol{o}) - count_{c_{m-1}}(\boldsymbol{o})$.

Let $\boldsymbol{o}$ be the outcome that $\textsf{MNW}_{\textsf{off}}$ chooses.
Since $\textsf{MNW}_{\textsf{off}}$ satisfies $PO$, and for every agent $i$ and round $t$, $v_i^t(c_{m}) > v_i^t(c_{m-1})$, then $count_{c_{m-1}}(\boldsymbol{o})=0$.
Let $q(\boldsymbol{o}) = \floor{\frac{u_2(\boldsymbol{o})}{m-1}}$, and $r(\boldsymbol{o}) = u_2(\boldsymbol{o}) \pmod{(m-1)}$. Clearly, $count_{c_m}(\boldsymbol{o}) \leq q(\boldsymbol{o})$. We show that $count_{c_m}(\boldsymbol{o}) = q(\boldsymbol{o})$. Indeed, assume by contradiction that ${count_{c_m}(\boldsymbol{o})} < q(\boldsymbol{o})$, 

let $\overline{\boldsymbol{o}}$ be an outcome, such that $$\overline{{o}}^t =  \begin{cases} 
        c_{m}, &  t < 1 + q(\boldsymbol{o}) \\
        c_{(1 + r(\boldsymbol{o}) + \mathbbm{1}_{r(\boldsymbol{o}) = m-2})}, & t = 1 + q(\boldsymbol{o}) \\  
      c_{1}, & t > 1 + q(\boldsymbol{o})
   \end{cases}.$$
That is, ${count_{c_m}(\overline{\boldsymbol{o}})} =  q(\boldsymbol{o}) + \mathbbm{1}_{r(\boldsymbol{o}) = m-2}$, and $count_{c_{m-1}}(\overline{\boldsymbol{o}}) = 0$ (since $1 + r(\boldsymbol{o}) + \mathbbm{1}_{r(\boldsymbol{o}) = m-2} \not = m-1$).
Note that $u_2(\overline{\boldsymbol{o}}) = \sum_{t=1}^{T} v^t_2(\overline{\boldsymbol{o}}^t) 
= v^t_2(c_{m}) * q(\boldsymbol{o}) + v^t_2(c_{(r(\boldsymbol{o}) + 1 + \mathbbm{1}_{r(\boldsymbol{o}) = m-2})}) + v^t_2(c_{1})(T - q(\boldsymbol{o}) - 1) 
= q(\boldsymbol{o}) * (m-1) + r(\boldsymbol{o}) + \mathbbm{1}_{r(\boldsymbol{o}) = m-2} 
= u_2(\boldsymbol{o}) + \mathbbm{1}_{r(\boldsymbol{o}) = m-2} \geq u_2(\boldsymbol{o})$. Moreover, since 
$u_1(\boldsymbol{o}) + u_2(\boldsymbol{o}) = T(m-1) + {count_{c_m}(\boldsymbol{o})}$ and 
$u_1(\overline{\boldsymbol{o}}) + u_2(\overline{\boldsymbol{o}}) = T(m-1) + {count_{c_m}(\overline{\boldsymbol{o}})}$
it follows that $u_1(\overline{\boldsymbol{o}}) = T(m-1) + {count_{c_m}(\overline{\boldsymbol{o}})} - u_2(\overline{\boldsymbol{o}}) = 
T(m-1) + {count_{c_m}(\overline{\boldsymbol{o}})} - u_2(\boldsymbol{o}) - \mathbbm{1}_{r(\boldsymbol{o}) = m-2} =
T(m-1) + q(\boldsymbol{o}) - u_2(\boldsymbol{o}) = 
T(m-1) + q(\boldsymbol{o}) - (T(m-1) + {count_{c_m}(\boldsymbol{o})} - u_1(\boldsymbol{o})) =
 q(\boldsymbol{o}) - {count_{c_m}(\boldsymbol{o})} + u_1(\boldsymbol{o})$.
Since we assume that $q(\boldsymbol{o}) > {count_{c_m}(\boldsymbol{o})}$ it follow that $u_1(\overline{\boldsymbol{o}}) > u_1(\boldsymbol{o})$, which is a contradiction to the fact that $\textsf{MNW}_{\textsf{off}}$ satisfies $PO$. Overall, we get that ${count_{c_m}(\boldsymbol{o})} = q(\boldsymbol{o})$, and thus  $u_1(\boldsymbol{o}) = q(\boldsymbol{o}) + {T(m-1)} - u_2(\boldsymbol{o})$.

We now show that $\textsf{MNW}_{\textsf{off}}$ chooses the outcome $\boldsymbol{o} = (c_m,\ldots,c_m)$. 
Clearly, $u_2((c_m,\ldots,c_m)) = T(n+1)$ and $\prod_{i \in N} u_i((c_m,\ldots,c_m)) = T^{n}{(n+1)}^{n-1}$.
Let $f$ be a function as defined in \autoref{lemma:mnw_not_m-5_app}. Since $m=n+2$, $\floor{\frac{u_2(\boldsymbol{o}))}{n+1}}(n+1) \leq u_2(\boldsymbol{o}) < (\floor{\frac{u_2(\boldsymbol{o}))}{n+1}} + 1)(n+1)$.
That is, $f(u_2(\boldsymbol{o})) = (\floor{\frac{u_2(\boldsymbol{o}))}{n+1}} + T(n+1) - u_2(\boldsymbol{o})){(u_2(\boldsymbol{o}))}^{n-1} = u_1(\boldsymbol{o}){(u_2(\boldsymbol{o}))}^{n-1} = \prod_{i \in N} u_i(\boldsymbol{o})$.
In addition, $f(u_2((c_m,\ldots,c_m))) = f(T(n+1)) = T^{n}{(n+1)}^{n-1}$.
Recall that $\textsf{MNW}_{\textsf{off}}$ chooses an outcome $\boldsymbol{o}$ such that $\boldsymbol{o} \in \argmax_{\boldsymbol{o'}} \prod_{i \in N} u_i(\boldsymbol{o'})$. That is, $\prod_{i \in N} u_i(\boldsymbol{o}) \geq \prod_{i \in N} u_i((c_m,\ldots,c_m))$. Therefore, $f(u_2(\boldsymbol{o})) = \prod_{i \in N} u_i(\boldsymbol{o}) \geq \prod_{i \in N} u_i((c_m,\ldots,c_m)) = f(u_2((c_m,\ldots,c_m))) = f(T(n+1))$. Since according to \autoref{lemma:mnw_not_m-5_app}, $f^{max} = \set{T(n+1)}$, 
$u_2(\boldsymbol{o}) = T(n+1)$. That is, $\textsf{MNW}_{\textsf{off}}$ chooses $\boldsymbol{o} = (c_m,\ldots,c_m)$.

Finally, if $T = \frac{n(n-1)}{2}$ then $u_1(\boldsymbol{o}) = T$, and $Prop_1  = \frac{T(m-1)}{n}$. Since $m = n+2$, $Prop_1 - u_1(\boldsymbol{o}) = \frac{T}{n} = \frac{n(n-1)}{2n} = \frac{(n-1)}{2} = \frac{(m-3)}{2}$. That is, $\textsf{MNW}_{\textsf{off}}$ does not guarantee an additive approximation of $Prop$ that is better than $\frac{m-3}{2}$.
\end{proof}

We now analyze $Prop1$. With Borda valuations, $\textsf{LMin}_{\textsf{off}}$ guarantees a $1$-additive approximation of $Prop$, and this entails that $\textsf{LMin}_{\textsf{off}}$ satisfies $Prop1$, as we show in \autoref{theorem:1_add_entails_prop1}.

We note that \citeauthor{conitzer2017fair}~\cite{conitzer2017fair} have raised an open question: is there a mechanism that satisfies $PO$, $Prop1$, and $RRS$ simultaneously? We partially solve this question- if we restrict the valuations to Borda valuations, then $\textsf{LMin}_{\textsf{off}}$ is such a mechanism, since we show that it satisfies $PO$, $Prop1$, and $RRS$.

Next, we consider $MPP$, and the negative results for $\textsf{MNW}_{\textsf{off}}$ and $\textsf{RR}_{\textsf{off}}$ still hold. Indeed, \autoref{exmp:rr_dont_find_prop} uses Borda valuations, and it shows that $\textsf{MNW}_{\textsf{off}}$ and $\textsf{RR}_{\textsf{off}}$ do not satisfy $MPP$.
In addition, $\textsf{MNW}_{\textsf{off}}$ still does not satisfy $RRS$.
\begin{theoremrep}
\label{theorem:mnw_rr_not_rrs_mpp}
Even With Borda valuations, $\textsf{MNW}_{\textsf{off}}$ does not satisfy $RRS$.
\end{theoremrep}
\begin{proof}
With Borda valuations, for any agent $i$, $RRS_i = \sum_{1 \leq t \leq \floor{T/n}}\boldsymbol{cMax_i}(t \cdot n) = \sum_{1 \leq t \leq \floor{T/n}} m-1 = \floor{\frac{T}{n}}(m-1)$. In addition, $Prop_i = \frac{1}{n}\sum_{t=1}^T v^t_i({cMax}^t_i) =  \frac{1}{n}\sum_{t=1}^T m-1 = {\frac{T}{n}}(m-1)$. Therefore, if $T \pmod n = 0$,  $RRS_i = Prop_i$. Recall the setting in the proof of \autoref{theorem:mnw_not_m-5_app}. The theorem shows that $\textsf{MNW}_{\textsf{off}}$ chooses an outcome, $\boldsymbol{o}$, that does not satisfy $Prop$, for any $n > 2$. If  $T=n$, $RRS_i = Prop_i$, and thus  $\boldsymbol{o}$ does not satisfy $RRS$. 
\end{proof}

Finally, we note that $\textsf{RR}_{\textsf{off}}$ still does not satisfy $PO$. 
\begin{theoremrep}
With Borda valuations, $\textsf{RR}_{\textsf{off}}$ does not satisfy $PO$. 
\end{theoremrep}
\begin{proof}
    Let $m = 4, n = 2, T = 2$ and $$V^1 = V^2 =  \begin{pmatrix}
3 & 2 & 1 & 0\\
0 & 2 & 1 & 3\\
\end{pmatrix}.$$
$\textsf{RR}_{\textsf{off}}$ chooses $\boldsymbol{o} = (c_1,c_4)$ or $\boldsymbol{o} = (c_4,c_1)$. That is, $u(\boldsymbol{o}) = (3,3)$. However, there is an outcome $\boldsymbol{o'} = (c_2,c_2)$ with $u(\boldsymbol{o'}) = (4,4)$.
\end{proof}

Overall, $\textsf{LMin}_{\textsf{off}}$ with Borda valuations is the ``fairest'' mechanism among the mechanisms we study, since it is the only one that satisfies $PO$, $Prop1$, $RRS$, $MPP$, and guarantees a $1$-additive approximation of $Prop$.

\section{Online Setting}
\label{sec:online}
Clearly, all the mechanisms we study do not satisfy the $PO$ efficiency property in the online setting. Moreover, we show that it is impossible to satisfy $Prop1$ or $\alpha$-$RRS$ for any constant $\alpha$.
\begin{theoremrep}
\label{theorm:not_rrs}
There is no online mechanism that satisfies $\alpha$-$RRS$, for any constant $\alpha$.
\end{theoremrep}
\begin{proof}
Assume by contradiction that there is an online mechanism that satisfies $\alpha$-$RRS$. Let $m = 2,n = 2,T = 3$, and $$V^1 =  \begin{pmatrix}
1 & 0\\
0 & 1
\end{pmatrix},X^2 = \begin{pmatrix}
\frac{2}{\alpha} & 0\\
0 & \frac{\alpha}{2}
\end{pmatrix},Y^2 = \begin{pmatrix}
\frac{\alpha}{2} & 0\\
0 & \frac{2}{\alpha}
\end{pmatrix}$$ $$X^3 = \begin{pmatrix}
\frac{4}{\alpha^2} & 0\\
0 & \frac{4}{\alpha^2}
\end{pmatrix},Y^3 = \begin{pmatrix}
0 & 0\\
0 & 0
\end{pmatrix}.$$
We set $V^2$ and $V^3$ according to the decisions of the online mechanism. Specifically, if $o^1 = c_1$ we set $V^2 = X^2$. Otherwise, $V^2 = Y^2$. Next, if $o^1 \not= o^2$ we set $V^3 = X^3$. Otherwise, $V^3 = Y^3$.


Note that for any agent $i$, $RRS_i \geq \frac{\alpha}{2}$ and thus $\alpha \cdot RRS_i > 0$. Since $\boldsymbol{o}$ satisfy $\alpha$-$RRS$ it must be the case that $o^1 \not= o^2$. Otherwise, $V^3 = Y^3$ and thus either $u_1(\boldsymbol{o}) = 0$ or $u_2(\boldsymbol{o}) = 0$, which is a contradiction. Therefore, $o^1 \not= o^2$ and $V^3 = X^3$. In addition, since $0 < \alpha \leq 1$, then $\frac{\alpha}{2} < 1 <  \frac{2}{\alpha} < \frac{4}{\alpha^2}$, and thus $\boldsymbol{cMax_i}$ equals $(1,\frac{2}{\alpha},\frac{4}{\alpha^2})$ or $(\frac{\alpha}{2},1,\frac{4}{\alpha^2})$, for any agent $i$. That is, $RRS_i \in \set{\frac{2}{\alpha},1}$, and thus $\alpha$-$RRS_i \in \set{2,\alpha}$, for any agent $i$.

Now, since $o^1 \not= o^2$, there are only $4$ possible outcomes:
\begin{enumerate}
    \item If $\boldsymbol{o} = (c_1,c_2,c_1)$, $V^2 = X^2$. Thus, $\alpha \cdot RRS_1 = 2$ and $\alpha \cdot RRS_2 = \alpha$. However, $u(\boldsymbol{o}) = (1 + \frac{4}{\alpha^2},\frac{\alpha}{2})$. That is, $u_2(\boldsymbol{o}) < \alpha \cdot RRS_2$.

    \item If $\boldsymbol{o} = (c_2,c_1,c_2)$, $V^2 = Y^2$. Thus, $\alpha \cdot RRS_1 = \alpha$ and $\alpha \cdot RRS_2 = 2$. However, $u(\boldsymbol{o}) = (\frac{\alpha}{2},1 + \frac{4}{\alpha^2})$. That is, $u_1(\boldsymbol{o}) < \alpha \cdot RRS_1$.

    \item If $\boldsymbol{o} = (c_1,c_2,c_2)$, $V^2 = X^2$. Thus,  $\alpha \cdot RRS_1 = 2$ and $\alpha \cdot RRS_2 = \alpha$. However, $u(\boldsymbol{o}) = (1 ,\frac{\alpha}{2} + \frac{4}{\alpha^2})$. That is, $u_1(\boldsymbol{o}) < \alpha \cdot RRS_1$.
    
    \item if $\boldsymbol{o} = (c_2,c_1,c_1)$, $V^2 = Y^2$. Thus, $\alpha \cdot RRS_1 = \alpha$ and $\alpha \cdot RRS_2 = 2$. However, $u(\boldsymbol{o}) = (\frac{\alpha}{2} + \frac{4}{\alpha^2},1)$. That is, $u_2(\boldsymbol{o}) < \alpha \cdot RRS_2$.

\end{enumerate}

Therefore, there is no online mechanism that satisfies $\alpha$-$RRS$.
\end{proof}

Next, we show that it is impossible to satisfy $Prop1$.
Let $Prop^t_i$ be the $Prop$ value of agent $i$ up to round $t$,  i.e., $Prop^t_i = \sum_{k = 1}^{k = t} \frac{v^k_i(cMax^k_i)}{n}$.
Let ${vProp1}^t_i = \argmax_{k \in \set{1,\ldots,t}} (v_i^k(cMax^k_i) - v_i^k(o^k))$. It is thus possible to rephrase the definition of $Prop1$: an outcome $\boldsymbol{o}$ satisfies $Prop1$ if for every agent $i$, $u_i(\boldsymbol{o}) + {vProp1}^T_i \geq Prop^T_i$. 
Now, let  ${dProp1}^t_i = u_i(\boldsymbol{o}^t) + {vProp1}^t_i - Prop^t_i$. That is, if ${dProp1}^t_i \geq 0$ for any $i$, then $\boldsymbol{o}^t$ satisfies $Prop1$ up to round $t$, and if ${dProp1}^t_i < 0$ for an agent $i$, then $Prop1$ is not satisfied up to round $t$. Therefore, intuitively, ${dProp1}^t_i$ represents the degree in which $\boldsymbol{o}^t$ satisfies $Prop1$. 

\begin{theoremrep}
\label{theorm:not_prop1}
There is no online mechanism that satisfies $Prop1$.
\end{theoremrep}
\begin{proofsketch}
Assume by contradiction that there is an online mechanism that satisfies $Prop1$.
We build a scenario with two agents, in which for each even round $t$, the mechanism must choose an outcome such that $dProp^t$ of one of the agents is the same as her $dProp^{t-2}$, but $dProp^t$ of the other agent is smaller than her $dProp^{t-2}$ by a constant factor. Therefore, after a constant number of rounds, one of the agents has a negative $dProp^{t}$, which entails that the mechanism does not satisfy $Prop1$.
\end{proofsketch}
\begin{proof}

We first show that an online mechanism that satisfies $Prop1$ for a given $T$, must also satisfy $Prop1$ for any $t < T$.

\begin{lemma}
\label{lemma:prop1_prefix}
If an online mechanism satisfies $Prop1$, then for any agent $i$ and round $t$, $u_i(\boldsymbol{o}^{t}) + {vProp1}^t_i \geq Prop^t_i.$
\end{lemma}
\begin{nestedproof}
    Assume by contradiction there is an online mechanism that satisfies $Prop1$, and there are an agent $i$ and a round $t$ in which $u_i(\boldsymbol{o}^{t}) + vProp1^t_i <  Prop^t_i$. Consider an instance with the same $m$,$n$, $T$, and the same valuations up to round $t$. For any round $\overline{t} > t$, let $v^{\overline{t}}_i(c) = 0$, for any candidate $c$. Given the new instance, let $\overline{\boldsymbol{o}}$ be the mechanism's outcome, and $vProp1^t_i$ and $Prop^t_i$ are now defined for the new instance.
    Clearly, $u_i(\overline{\boldsymbol{o}}^{t}) + {vProp1}^t_i < Prop^t_i$, and since $v^{\overline{t}}_i(c) = 0$ for any candidate $c$, $u_i(\overline{\boldsymbol{o}}^{T}) + {vProp1}^T_i < Prop^T_i$. That is, $\overline{\boldsymbol{o}}$ does not satisfy $Prop1$, which is a contradiction.
\end{nestedproof}
We show that, for two agents, the relation between ${dProp1}^t_i$ and ${vProp1}^t_i$ does not depend on $t$.
\begin{lemma}
\label{lemma:prop1_dprop_lower_bound}
Consider an outcome of an online mechanism $\boldsymbol{o}$ that satisfies $Prop1$, and let $n = 2$.
For any agent $i$, and for any round $t$ such that $t \leq T - 3$, $8 \cdot {dProp1}^t_i \geq {vProp1}^t_i.$
\end{lemma}
\begin{nestedproof}
Let $n = 2$. Assume by contradiction that there is an online mechanism that satisfies $Prop1$, and there is a round $t$, $t \leq T - 3$, and an agent $i$, such that $8 \cdot {dProp1}^t_i < {vProp1}^t_i$. Since the mechanism satisfies $Prop1$, then ${dProp1}^t_i \geq 0$. If ${dProp1}^t_i > 0$, there is an $\epsilon > 1$, ${dProp1}^t_i > \epsilon - 1$, such that $8\epsilon^3 {dProp1}^t_i < {vProp1}^t_i$. If ${dProp1}^t_i = 0$, there is an $\epsilon > 1$ such that $8\epsilon^3 (\epsilon - 1) < {vProp1}^t_i$. Overall, there is an $\epsilon > 1$ such that $8\epsilon^3 \max\set{{dProp1}^t_i,\epsilon - 1} < {vProp1}^t_i$.
    
Now, consider an instance with the same $m$, $n$, $T$, and the same valuations up to round $t$. For the agent $i$ and a round $\overline{t} > t$, let $v^{\overline{t}}_i(c_1) = 2\epsilon{(\epsilon +1)}^{\overline{t} - t - 1} \max\set{{dProp1}^t_i,\epsilon - 1}$, and $v^{\overline{t}}_i(c) = 0$ for any $c \neq c_1$. For the other agent $j \neq i$ and a round $\overline{t} > t$, let $v^{\overline{t}}_j(c_2) = 4( Prop^t_j + 1)$, and $v^{\overline{t}}_j(c) = 0$, for any $c \not = c_2$. Given the new instance, $vProp1^t_i$, $Prop^t_i$, and $dProp^t_i$ are now defined for the new instance(for all the agents). Clearly, it is still holds that $8\epsilon^3 \max\set{{dProp1}^t_i,\epsilon - 1} < {vProp1}^t_i$.

We first show that $vProp1^{\overline{t}}_i = vProp1^{t}_i$, for any round $\overline{t}$ such that  $t + 3 \geq \overline{t} \geq t$. Since $v^{\overline{t}}_i(c_1) > 0$, but $v^{\overline{t}}_i(c) = 0$ for any $c \not = c_1$, then $cMax^{\overline{t}}_i = c_1$. Therefore,  $vProp1^{t+1}_i = \max\set{vProp1^{t}_i,v^{t + 1}_i(c_1)}$, $vProp1^{t+2}_i = \max\set{vProp1^{t}_i,v^{t + 1}_i(c_1),v^{t + 2}_i(c_1)}$, and $vProp1^{t+3}_i = \max\set{vProp1^{t}_i,v^{t + 1}_i(c_1),v^{t + 2}_i(c_1),v^{t + 3}_i(c_1)}$. In addition, for $\overline{t} \in \set{t+1,t+2,t+3}$, 
\begin{equation*}
    \begin{split}
   v^{\overline{t}}_i(c_1)  
    &= 2\epsilon{(\epsilon +1)}^{\overline{t} - t - 1} {\max\set{{dProp1}^t_i,\epsilon - 1}} \\
    &\leq 2\epsilon{(\epsilon +1)}^{2} {\max\set{{dProp1}^t_i,\epsilon - 1}} \\
    &\leq 2\epsilon{(\epsilon + \epsilon)}^{2} {\max\set{{dProp1}^t_i,\epsilon - 1}} \\
    &= 8{\epsilon}^{3} {\max\set{{dProp1}^t_i,\epsilon - 1}} \\
    &< {vProp1}^t_i 
    \end{split}
\end{equation*}
That is, $vProp1^{t}_i = vProp1^{t + 1}_i = vProp1^{t + 2}_i =  vProp1^{t + 3}_i$.

We now show that $o^{\overline{t}} = c_1$, for $\overline{t} \in \set{t+1,t+2,t+3}$. Note that $Prop^{{\overline{t}}}_i 
    = \frac{1}{2} \sum_{k = 1}^{k = \overline{t}} v^k_j(cMax^k_i)
    = \frac{1}{2} \sum_{k = 1}^{k = \overline{t} - 1} v^k_i(cMax^k_i) + \frac{1}{2}v^{\overline{t}}_i(cMax^{\overline{t}}_i)
    = Prop^{\overline{t} - 1}_i + \frac{1}{2}v^{\overline{t}}_i(cMax^{\overline{t}}_i)
    = Prop^{\overline{t} - 1}_i + \frac{1}{2}v^{\overline{t}}_i(c_1)$.
    In addition, ${dProp1}^{{\overline{t}}}_i 
    = u_i(\boldsymbol{o}^{{\overline{t}}}) + {vProp1}^{{\overline{t}}}_i -  Prop^{{\overline{t}}}_i
    = v^{\overline{t}}_i(o^{\overline{t}}) + u_i(\boldsymbol{o}^{\overline{t} - 1}) + {vProp1}^{\overline{t} - 1}_i -  Prop^{\overline{t} - 1}_i - \frac{1}{2}v^{\overline{t}}_i(c_1) 
    = v^{\overline{t}}_i(o^{\overline{t}}) + {dProp1}^{{\overline{t} - 1}}_i - \frac{1}{2}v^{\overline{t}}_i(c_1)$.

    Clearly when $\overline{t} = t$, then ${dProp1}^{\overline{t}}_i \leq {\max\set{{dProp1}^{\overline{t}}_i,\epsilon - 1}} = {(\epsilon +1)}^{\overline{t} - t} {\max\set{{dProp1}^t_i,\epsilon - 1}}$.
    We show that if  ${dProp1}^{\overline{t}}_i \leq {(\epsilon +1)}^{\overline{t} - t} {\max\set{{dProp1}^t_i,\epsilon - 1}}$ then also ${dProp1}^{\overline{t} + 1}_i \leq {(\epsilon +1)}^{\overline{t} + 1 - t} {\max\set{{dProp1}^t_i,\epsilon - 1}}$ for some $t + 3 > \overline{t} \geq t$.
    Assume that  ${dProp1}^{\overline{t}}_i \leq {(\epsilon +1)}^{\overline{t} - t} {\max\set{{dProp1}^t_i,\epsilon - 1}}$ for some $t + 3 > \overline{t} \geq t$.

    Assume by contradiction that $o^{\overline{t} + 1} \not= c_1$. 
    Since $o^{t + 1} \not= c_1$, then $v^{\overline{t} + 1}_i(o^{\overline{t} + 1}) = 0$.

    That is, \begin{equation*}
        \begin{split}
       {dProp1}^{{\overline{t}} + 1}_i  
        &= {dProp1}^{{\overline{t}}}_i - \frac{1}{2} v^{\overline{t} + 1}_i(c_1) \\
        &\leq {(\epsilon  +1)}^{\overline{t} - t} {\max\set{{dProp1}^t_i,\epsilon - 1}} - \frac{1}{2} v^{\overline{t} + 1}_i(c_1) \\
        &= {(\epsilon  +1)}^{\overline{t} - t} {\max\set{{dProp1}^t_i,\epsilon - 1}} \\
        &-\frac{1}{2}2 \epsilon{(\epsilon + 1)}^{\overline{t} - t} {\max\set{{dProp1}^t_i,\epsilon - 1}} \\
        &= {(\epsilon + 1)}^{\overline{t} - t} {\max\set{{dProp1}^t_i,\epsilon - 1}}(1 - \epsilon) 
        \end{split}
    \end{equation*}
    However, since $\epsilon > 1$, then $(1 - \epsilon) < 0$ and ${(\epsilon + 1)}^{\overline{t} - t} {\max\set{{dProp1}^t_i,\epsilon - 1}} > 0$. That is, ${dProp1}^{{\overline{t}} + 1}_i < 0$, and thus   $u_i(\boldsymbol{o}^{{\overline{t}}+1}) + {vProp1}^{\overline{t} + 1}_i <  Prop^{{\overline{t}} + 1}_i$, which is a contradiction to the fact that the mechanism satisfies $Prop1$ (according to \autoref{lemma:prop1_prefix}).
    Hence, $o^{{\overline{t}} + 1} = c_1$. We now can now find an upper bound on ${dProp1}^{\overline{t} + 1}_i$:

    \begin{equation*}
        \begin{split}
          {dProp1}^{\overline{t} + 1}_i   
            &= v^{\overline{t} + 1}_i(o^{\overline{t} + 1}) + {dProp1}^{{\overline{t}}}_i - \frac{1}{2}v^{\overline{t} + 1}_i(c_1) \\
            &= \frac{1}{2}v^{\overline{t} + 1}_i(c_1) + {dProp1}^{{\overline{t}}}_i \\
            &= \frac{1}{2}2\epsilon{(\epsilon +1)}^{\overline{t} - t} {\max\set{{dProp1}^t_i,\epsilon - 1}} + {dProp1}^{{\overline{t}}}_i \\
             &= {(\epsilon +1)}^{\overline{t} - t} {\max\set{{dProp1}^t_i,\epsilon - 1}}\epsilon + {dProp1}^{{\overline{t}}}_i \\
            &\leq {(\epsilon +1)}^{\overline{t} - t} {\max\set{{dProp1}^t_i,\epsilon - 1}}(\epsilon + 1) \\
        \end{split}
    \end{equation*} 

    That is, ${dProp1}^{\overline{t} + 1}_i  \leq {(\epsilon +1)}^{\overline{t} + 1 - t} {\max\set{{dProp1}^t_i,\epsilon - 1}}$. Therefore, ${dProp1}^{\overline{t}}_i \leq {(\epsilon +1)}^{\overline{t} - t} {\max\set{{dProp1}^t_i,\epsilon - 1}}$, and $o^{\overline{t}} = c_1$ for any $t + 3 \geq \overline{t} > t$.

    We will now show that $Prop1$ is not satisfied for agent $j$ such that $j \neq i$. We will do that by showing that ${dProp1}^{t + 3}_{j}  < 0$. We thus need to find bounds for $u_j(\boldsymbol{o}^{t + 3}),{vProp1}^{t + 3}_j$ and $Prop^{t + 3}_j$.

    We will start with $u_j(\boldsymbol{o}^{t + 3})$. Since  $o^{\overline{t}} = c_1$ for any $t + 3 \geq \overline{t} > t$, then $v^{\overline{t}}_j(o^{\overline{t}}) = 0$. In addition, by definition, $u_j(\boldsymbol{o}^{t + 3}) = \sum_{k = 1}^{k = t + 3} v^k_j({o}^{k})$. Therfore, $u_j(\boldsymbol{o}^{t + 3}) = \sum_{k = 1}^{k = t} v^k_j({o}^{k}) + \sum_{k = t + 1}^{k = t + 3} v^k_j({o}^{k}) = u_j(\boldsymbol{o}^{t})$. In addition,recall that $v^k_j({o}^{k}) \leq v^k_j(cMax^k_j)$. We thus get that, $u_j(\boldsymbol{o}^{t}) = \sum_{k = 1}^{k = t} v^k_j({o}^{k}) \leq 2 \frac{1}{2} \sum_{k = 1}^{k = t} v^k_j(cMax^k_j) \leq  2 \cdot Prop^{t}_j$. That is,  $u_j(\boldsymbol{o}^{t}) \leq 2 \cdot Prop^{t}_j$. Therfore, $u_j(\boldsymbol{o}^{t + 3}) \leq 2 \cdot Prop^{t}_j$ 

    By definition, ${vProp1}^{t + 3}_j = \argmax_{k \in \set{1,\ldots,t + 3}} (v_j^k(cMax^k_j) - v_j^k(o^k))$. That is, ${vProp1}^{t + 3}_j = \max\set{{vProp1}^{t}_j,v_j^{t+1}(cMax^{t+1}_j) - v_j^{t+1}(o^{t+1}), v_j^{t+2}(cMax^{t+2}_j) - v_j^{t+2}(o^{t+2}), v_j^{t+3}(cMax^{t+3}_j) - v_j^{t+3}(o^{t+3})}$. Recall that $v^{\overline{t}}_j(c_2) = 4( Prop^t_j + 1)$, and $v^{\overline{t}}_j(c) = 0$, for any other $c \neq c_2$. In addition,  $o^{\overline{t}} = c_1$ for any $t + 3 \geq \overline{t} > t$. We thus get that, ${vProp1}^{t + 3}_j = \max\set{{vProp1}^{t}_j,4( Prop^t_j + 1)}$. Clearly, ${vProp1}^{t}_j = \argmax_{k \in \set{1,\ldots,t + 3}} (v_j^k(cMax^k_j) - v_j^k(o^k)) \leq \sum_{k = 1}^{k = t} v^k_j(cMax^k_j) = 2 \frac{1}{2}\sum_{k = 1}^{k = t} v^k_j(cMax^k_j) = 2Prop^t_j < 4( Prop^t_j + 1)$. That is, ${vProp1}^{t}_j <  4( Prop^t_j + 1)$. We thus get that ${vProp1}^{t + 3}_j = 4(Prop^t_j + 1)$.
    
    By definition, $Prop^{t + 3}_j = \frac{1}{2} \sum_{k = 1}^{k = t + 3} v^k_j(cMax^k_j) = Prop^t_j + (\frac{3}{2})v^{t+1}_j(cMax^{t+1}_j) = Prop^t_j + (\frac{3}{2})4( Prop^t_j + 1) = Prop^t_j + 6Prop^t_j + 6$. That is,  $Prop^{t + 3}_j =  7Prop^t_j  + 6$.

    Overall, ${dProp1}^{t + 3}_{j} = u_j(\boldsymbol{o}^{t + 3}) + {vProp1}^{t + 3}_j -  Prop^{t + 3}_j = u_j(\boldsymbol{o}^{t + 3}) + 4(Prop^t_j + 1) - 7Prop^t_j - 6 = u_j(\boldsymbol{o}^{t + 3}) -3Prop^t_j -2 \leq 2 \cdot Prop^{t}_j -3\cdot Prop^t_j -2 = - Prop^t_j -2 < 0$.
    That is,  ${dProp1}^{t + 3}_{j}  < 0$, a contradiction according to \autoref{lemma:prop1_prefix}. 

    Hence, for any agent $i$, and for any round $t$, such that $t \leq T - 3$, $8 {dProp1}^t_i \geq {vProp1}^t_i.$
    
\end{nestedproof}

We are now ready to prove that no online mechanism satisfies $Prop1$.

Assume by contradiction that there is an online mechanism that satisfies $Prop1$.
Let $\epsilon = \frac{1}{8}$.
Let $T = 2k + 3$, where $k \in N$ and $k > \frac{4}{\epsilon}$.
In addition, let $m = 2$, $n = 2$, and 
    $$V^1 =  \begin{pmatrix}
1 & 0\\
0 & 1
\end{pmatrix},X= \begin{pmatrix}
1 & 0\\
0 & 1 -\epsilon
\end{pmatrix},Y= \begin{pmatrix}
1-\epsilon & 0\\
0 & 1
\end{pmatrix}.$$
For every $t \leq T$, let 
$$V^{t} =  \begin{cases}
    V^1,&\text{if }  t \in \set{2x + 1\mid x \in \mathbb{N}}\\
    X,&\text{if }  t \in \set{2x\mid x \in \mathbb{N}} \land o^{t-1} = c_1\\
    Y,&\text{if }  t \in \set{2x\mid x \in \mathbb{N}} \land o^{t-1} = c_2
\end{cases}.$$

We first show that if $v^{2t}_i({o}^{2t}) = 1 - \epsilon$, then $v^{2t - 1}_i({o}^{2t - 1}) = 0$, for any round $2t$ and agent $i$.
Assume that $v^{2t}_1({o}^{2t}) = 1 - \epsilon$. Therefore, $V^{2t} = Y$, and $o^{2t-1} = c_2$. That is, $v^{2t - 1}_1({o}^{2t - 1}) = 0$.
Similarly, assume that $v^{2t}_2({o}^{2t}) = 1 - \epsilon$. Therefore, $V^{2t} = X$, and $o^{2t-1} = c_1$. That is, $v^{2t - 1}_2({o}^{2t - 1}) = 0$.

We now show that if $v^{2t}_i({o}^{2t}) = 1$, then $v^{2t - 1}_i({o}^{2t - 1}) = 1$, for any round $2t$ and agent $i$.
Assume that $v^{2t}_1({o}^{2t}) = 1$. Therefore, $V^{2t} \not= Y$, and $o^{2t-1} = c_1$. That is, $v^{2t - 1}_1({o}^{2t - 1}) = 1$.
Similarly, assume that $v^{2t}_2({o}^{2t}) = 1 - \epsilon$. Therefore, $V^{2t} \not= X$, and $o^{2t-1} = c_2$. That is, $v^{2t - 1}_2({o}^{2t - 1}) = 1$.
%
Overall, there are only $4$ possible valuations for any agent $i$, and pairs of round $2t - 1$ and $2t$:
\begin{enumerate}
    \item $v^{2t - 1}_i({o}^{2t - 1}) = 0$ and $v^{2t}_i({o}^{2t}) = 1 -\epsilon$.
    \item $v^{2t - 1}_i({o}^{2t - 1}) = 1$ and $v^{2t}_i({o}^{2t}) = 1$.
    \item $v^{2t - 1}_i({o}^{2t - 1}) = 0$ and $v^{2t}_i({o}^{2t}) = 0$.
    \item $v^{2t - 1}_i({o}^{2t - 1}) = 1$ and $v^{2t}_i({o}^{2t}) = 0$.
\end{enumerate}
Let ${CountPairs}^t_i(a,b)$ be the number of times agent $i$ received in pairs $a$ and $b$ in pairs of rounds $2r - 1 \leq t$ and $2r \leq t$. That is, ${CountPairs}^t_i(a,b) =  |\set{r \in [1,\frac{t}{2}] \cap \mathbb{N} \mid v^{2r - 1}_i({o}^{2r - 1}) = a \land  v^{2r}_i({o}^{2r}) = b}|$. We thus defines (1) ${{\epsilon}Count}^t_i = {CountPairs}^t_i(0,1-\epsilon)$. (2) ${Count11}^t_i = {CountPairs}^t_i(1,1)$. (3) ${Count00}^t_i = {CountPairs}^t_i(0,0)$. (4) ${Count10}^t_i = {CountPairs}^t_i(1,0)$.

We can now compute $u_i(\boldsymbol{o}^{2t})$, for any round $2t$. Clearly, each instance of case (1) increases $u_i(\boldsymbol{o}^{2t})$ by $1-\epsilon$. (2) increases $u_i(\boldsymbol{o}^{2t})$ by $2$. (3) does not effect $u_i(\boldsymbol{o}^{2t})$. (4) increases $u_i(\boldsymbol{o}^{2t})$ by $1$. Overall, $$u_i(\boldsymbol{o}^{2t}) = {{\epsilon}Count}^{2t}_i(1 - \epsilon) + 2 \cdot {Count11}^{2t}_i + {Count10}^t_i.$$ 

We now compute  $Prop^{2t}_i$, for any round $2t$. Clearly, $v^{2t-1}_i(cMax^{2t - 1}) = 1$. In cases (1) and (3) $v^{2t}_i(cMax^{2t}) = 1-\epsilon$, and in cases (2) and (4) $v^{2t}_i(cMax^{2t}) = 1$. We thus get that
$Prop^{2t}_i = \frac{1}{2}(({{\epsilon}Count}^t_i + {Count11}^t_i + {Count00}^t_i + {Count10}^t_i) + (1 - \epsilon)({{\epsilon}Count}^t_i + {Count00}^t_i) + ({Count11}^t_i + {Count10}^t_i))$. That is, 
$Prop^{2t}_i = ({{\epsilon}Count}^{2t}_i + {Count00}^{2t}_i)(1 - \frac{\epsilon}{2}) + {Count11}^{2t}_i + {Count10}^{2t}_i$.

As for ${dProp1}^{2t}_i$, by definition, ${dProp1}^{2t}_i = u_i(\boldsymbol{o}^{2t}) + {vProp1}^{2t}_i -  Prop^{2t}_i$. That is, ${dProp1}^{2t}_i = {{\epsilon}Count}^{2t}_i(1 - \epsilon) + 2 \cdot {Count11}^{2t}_i + {Count10}^{2t}_i +  {vProp1}^{2t}_i - (({{\epsilon}Count}^{2t}_i + {Count00}^{2t}_i)(1 - \frac{\epsilon}{2}) + {Count11}^{2t}_i + {Count10}^{2t}_i)$. That is, ${dProp1}^{2t}_i = {vProp1}^{2t}_i - {{\epsilon}Count}^{2t}_i\frac{\epsilon}{2} - {Count00}^{2t}_i(1 - \frac{\epsilon}{2}) + {Count11}^{2t}_i$.

We claim that $o^{2t - 1} \not= o^{2t}$, for any $2t \leq T - 3$.
Assume that $o^{2r - 1} \not= o^{2r}$, for any $r \in \set{1,\ldots,t}$, for some $2t \leq T - 5$.
Clearly, the claim holds if $t = 0$ and $T \geq 5$.
Assume by contradiction that $o^{2{t} + 1} = o^{2{t} + 2}$.
Let $i$ be the agent such that $o^{2{t} + 1} \not = c_i$. Since $o^{2t + 1} \not=c_i$, then $v^{2t+1}_i(o^{2t + 1}) = v^{2t+2}_i(o^{2t + 2}) = 0$. Therefore, ${vProp1}^{2t + 2}_i = 1$.
Since $o^{2r - 1} \not= o^{2r}$, for any $r \in \set{1,\ldots,t}$, for some $2t \leq T - 5$, then ${Count11}^{2t + 2}_i = 0$ and ${Count00}^{2t + 2}_i = 1$.
That is, ${dProp1}^{2t + 2}_i = {vProp1}^{2t + 2}_i - {{\epsilon}Count}^{2t + 2}_i\frac{\epsilon}{2} -{Count00}^{2t + 2}_i(1 - \frac{\epsilon}{2}) + {Count11}^{2t + 2}_i = (1 - {{\epsilon}Count}^{2t + 2}_i)\frac{\epsilon}{2}$.
Now, according to \autoref{lemma:prop1_dprop_lower_bound} and the fact that the online mechanism satisfy $Prop1$,
$$8 \cdot {dProp1}^{2t + 2}_i \geq {vProp1}^{2t + 2}_i.$$
Since $\epsilon = \frac{1}{8}, {vProp1}^{2t + 2}_i = 1,$ and ${dProp1}^{2t + 2}_i = (1 - {{\epsilon}Count}^{2t + 2}_i)\frac{\epsilon}{2}$, we get that $(1 - {{\epsilon}Count}^{2t + 2}_i)\frac{1}{2} \geq 1$. Clearly, $(1 - {{\epsilon}Count}^{2t + 2}_i) \leq 1$. That is $\frac{1}{2} \geq (1 - {{\epsilon}Count}^{2t + 2}_i)\frac{1}{2} \geq 1$, a contradiction. Therefore, $o^{2{t} + 1} \not= o^{2{t} + 2}$.

Overall, since $o^{2t - 1} \not= o^{2t}$, for any $2t \leq T - 3$, ${Count00}^{2t}_i = {Count11}^{2t}_i = 0$, for any agent $i$ and for any round $2t \leq T - 3$.
Moreover, ${{\epsilon}Count}^{2t}_1 + {{\epsilon}Count}^{2t}_2 = t$, since in any pairs of round $2t \leq T - 3$ and $2t - 1 \leq T - 3$, one of the agents recives a utility of $1-\epsilon$, and there are exactly $t$ such pairs.

Recall that $T = 2k+3$, and consider ${dProp1}^{2k}_i$. By definition, $2k \leq T - 3$, and thus
${dProp1}^{2k}_i = {vProp1}^{2k}_i - {{\epsilon}Count}^{2k}_i\frac{\epsilon}{2} -{Count00}^{2k}_i(1 - \frac{\epsilon}{2}) + {Count11}^{2k}_i = {vProp1}^{2k}_i - {{\epsilon}Count}^{2k}_i\frac{\epsilon}{2}$.

Finally, consider a lower bound on ${dProp1}^{2k}_1 + {dProp1}^{2k}_2$.
${dProp1}^{2k}_1 + {dProp1}^{2k}_2 =  ({vProp1}^{2k}_1 - {{\epsilon}Count}^{2k}_1\frac{\epsilon}{2}) + ({vProp1}^{2k}_2 - {{\epsilon}Count}^{2k}_2\frac{\epsilon}{2}) = ({vProp1}^{2k}_1 + {vProp1}^{2k}_2) - \frac{\epsilon}{2}({{\epsilon}Count}^{2k}_1 + {{\epsilon}Count}^{2k}_2)$. Since ${{\epsilon}Count}^{2k}_1 + {{\epsilon}Count}^{2k}_2 = k$, then ${dProp1}^{2k}_1 + {dProp1}^{2k}_2 = ({vProp1}^{2k}_1 + {vProp1}^{2k}_2) - k\frac{\epsilon}{2}$. Clearly, $({vProp1}^{2k}_1 + {vProp1}^{2k}_2) \leq 2$, and thus ${dProp1}^{2k}_1 + {dProp1}^{2k}_2 \leq 2 - k\frac{\epsilon}{2}$.
Since $k > \frac{4}{\epsilon}$, then $2 - k\frac{\epsilon}{2} < 0$. That is, ${dProp1}^{2k}_1 + {dProp1}^{2k}_2 < 0$. Therefore, either ${dProp1}^{2k}_1 < 0$ or ${dProp1}^{2k}_2 < 0$, which is a contradiction to the fact that the mechanism satisfies $Prop1$ (according to \autoref{lemma:prop1_prefix}). Therefore, there is no online mechanism that satisfies $Prop1$.
\end{proof}

Note that the proofs of Theorems~\ref{theorm:not_rrs} and \ref{theorm:not_prop1} essentially use the setting of indivisible private goods. Therefore, they strengthen previous results on online fair division ~\cite{he2019achieving}. Overall, with no restriction on the valuations, there is no online mechanism that satisfies even the relaxed fairness properties (i.e., $RRS$ or $Prop1$).

\subsection{Borda Valuations}
We begin with analyzing $Prop$. Our main result here is that $\textsf{LMin}_{\textsf{on}}$ guarantees a $1$-additive approximation of $Prop$. To prove this, we first show a unique characteristic of Borda valuations.

\begin{lemmarep} 
\label{lem:borda_sum_m-1} With Borda valuations, given a round $t$, and given any vector of $n$ elements, $\boldsymbol{x}$, such that (1) every element of $\boldsymbol{x}$ is a non-negative integer, and (2) the sum of the elements is $m-1$, there exists a candidate $c \in C$ such that $v_i^t(c) \geq \boldsymbol{x}_i$ for every agent $i$.
\end{lemmarep}
\begin{proof} 
Let  $\boldsymbol{x}$ be a vector of $n$ elements, such that (1) every element of $\boldsymbol{x}$ is a non-negative integer, and (2) the sum of the elements is $m-1$. For each agent $i$, let $B_i$ be the set of candidates whose valuations are less than $\boldsymbol{x}_i$, i.e., $B_i = \{c \in C \mid v_i^t(c) < \boldsymbol{x}_i \}$. Let $G$ be a (possibly empty) set of candidates, $G = C \setminus (B_1 \bigcup \ldots \bigcup B_n)$. Clearly, $|G| \geq m - \sum |B_i|$. 
Since the agents use Borda valuations, then $|B_i| = \boldsymbol{x}_i$, and thus $|G| \geq m-\sum \boldsymbol{x}_i = m - (m-1) = 1$. That is,  
a candidate $c \in G$ exists such that $v_i^t(c) \geq \boldsymbol{x}_i$ for every agent $i$.
\end{proof}

Intuitively, \autoref{lem:borda_sum_m-1} shows that any mechanism can choose an outcome such that the sum of all the agents' valuations is at least $T(m-1)$. Moreover, a mechanism is able to distribute the valuations arbitrarily, and thus it can choose an outcome that guarantees a utility of $\floor{\frac{T(m-1)}{n}}$ for each agent. Therefore, $\textsf{LMin}_{\textsf{off}}$ must also guarantee a utility of $\floor{\frac{T(m-1)}{n}}$ for each agent, which means that is satisfies $1$-additive approximation of $Prop$. The $1$-additive approximation for the online setting is also based on \autoref{lem:borda_sum_m-1}. Intuitively, the lemma shows that for each round, the vector of elements $x$ can be chosen without knowing the  valuations of each agent. Therefore, the vector $x$ for each round can be chosen greedily, and it thus determines each $o^t$ in round $t$ such that the utility of $\boldsymbol{o}$ is at least $\floor{\frac{T(m-1)}{n}}$ for each agent.

Next, we show an important property of proportionality in the online setting, i.e., when the candidates of the chosen outcome are determined sequentially. The proof is essentially an extension of \autoref{lem:borda_sum_m-1} to the online setting.
Recall that with Borda valuations, $Prop_i = \frac{T(m-1)}{n}$ for every agent $i$. Let $Prop^t = \frac{t(m-1)}{n}$,  $qProp^t = \floor{{Prop}^t}$, and $rProp^t = (t(m-1)) \bmod n$.

\begin{lemmarep}
\label{lemma:lemma_addtive_main}
With Borda valuations, given a round $t$, if (1) for every agent $i$, $u_i(\boldsymbol{o}^{t-1}) \geq qProp^{t-1}$ and
(2) there are at least  $rProp^{t-1}$ agents with a utility that is strictly greater than $qProp^{t-1}$, then there exists a candidate $c$ such that (3) for every agent $i$, $u_i((o^1,\ldots,o^{t-1},c)) \geq qProp^{t}$ and (4) there are at least  $rProp^{t}$ agents with a utility that is strictly greater than $qProp^{t}$.
\end{lemmarep} 

\begin{proof}

To prove this lemma we need the next lemma, which shows a simple number theoretic property.
\begin{lemma} 
\label{lemma:postive_integer_properties}
Let $a, b, n$ be positive integers, and let $x$ be a real number. Then, the following statements are equivalent: (1) $\floor{\frac{a}{n}} + \floor{\frac{b}{n}} + x = \floor{\frac{a+b}{n}}$. (2) $(a \bmod n) + (b \bmod n) = xn + ((a+b) \bmod n)$. Furthermore, there exist such $a, b, n$ if and only if $x \in \set{0, 1}$. \end{lemma}
\begin{nestedproof}
Clearly, for any positive integer $z$, $\floor{\frac{z}{n}} \cdot n + (z \bmod n) = z$. Therefore, $\floor{\frac{z}{n}} = \frac{z - (z \bmod n)}{n}$. We can thus show that statements (1) and (2) are equivalent: $ \floor{\frac{a}{n}} + \floor{\frac{b}{n}} + x = \floor{\frac{a+b}{n}} \Longleftrightarrow \frac{a - (a \bmod n)}{n} + \frac{b - (b \bmod n)}{n} + x = \frac{(a+b) - ((a+b) \bmod n)}{n} \Longleftrightarrow (a \bmod n) + (b \bmod n) = xn + ((a+b) \bmod n) $. Now, consider the possible values of $x$:
\begin{enumerate}
    \item  Assume that $x \geq 2$. Then, $xn \geq 2n$, and thus $xn + ((a+b) \bmod n) \geq 2n$. 
    That is, $(a \bmod n) + (b \bmod n) \geq 2n$.
    However, since $(a \bmod n) < n$ and $(b \bmod n) < n$, it follows that $(a \bmod n) + (b \bmod n) < 2n$, which is a contradiction. Therefore, $x < 2$.
    \item Assume that $x \leq -1$. Then, $xn < -n$, and thus $0 > xn + n > xn + ((a+b) \bmod n) = (a \bmod n) + (b \bmod n)$. However, since $(a \bmod n) \geq 0$ and $(b \bmod n) \geq 0$, it follows that $(a \bmod n) + (b \bmod n) \geq 0$, which is a contradiction. Therefore, $x > -1$.
\end{enumerate}
 In addition, since $\floor{\frac{a+b}{n}}, \floor{\frac{a}{n}}$ and $\floor{\frac{b}{n}}$ are integers, and  $x = \floor{\frac{a+b}{n}} - \floor{\frac{a}{n}} - \floor{\frac{b}{n}}$, $x$ must also be an integer. That is, $x \in \set{0, 1}$.
\end{nestedproof}

We now prove \autoref{lemma:lemma_addtive_main}.
Given an outcome $\boldsymbol{o}^t$, let ${Q}^t = \set{i \mid u_i(\boldsymbol{o}^t) \geq {qProp}^t}$, ${Q}^t_> = \set{i \mid u_i(\boldsymbol{o}^t) \geq {qProp}^t + 1}$, and
$\overline{{Q}^{t}_>} = N \setminus {Q}^{t}_>$. Let $\overline{{sQ}^{t}_>}$ be some arbitrary subset of $\overline{{Q}^{t}_>}$ such that $|\overline{{sQ}^{t}_>}| = \min{(|\overline{{Q}^{t}_>}|, {rProp^1})}$, and let ${sN}^t$ be some arbitrary subset of $N$ such that $|{sN}^t| = {rProp^1} - |\overline{{sQ}^{t}_>}|$. Let $\boldsymbol{x}^t$ be a vector such that $\boldsymbol{x}^{t}_i = qProp^{1} + \mathbbm{1}_{i \in {sN}^t} + \mathbbm{1}_{i \in \overline{{sQ}^{t}_>}}$. Note that $\sum_i \boldsymbol{x}^{t-1}_i = \sum_i qProp^1 + \mathbbm{1}_{i \in {sN}^{t-1}} + \mathbbm{1}_{i \in \overline{{sQ}^{t-1}_>}} = n\cdot{qProp^1} + |{sN}^{t-1}| + |\overline{{sQ}^{t-1}_>}| = n\cdot{qProp^1} + rProp^1 = m-1$.

For a given round $t$, assume that conditions (1) and (2) hold. That is, ${Q}^{t-1} = C$, and $|{Q}^{t-1}_>| \geq {rProp}^{t-1}$. Then, according to \autoref{lem:borda_sum_m-1} there exists a candidate $c$ such that ${v({c})}^{t}_i \geq \boldsymbol{x}^{t-1}_i$, for every agent $i$. We will now show that if $o^t = c$, then ${Q}^t = C$ and $|{Q}^{t}_>| \geq rProp^t$.

\newcommand{\fakeimage}{{\fboxsep=-\fboxrule\fbox{\rule{0pt}{3cm}\hspace{4cm}}}}

Indeed, if $o^t = c$ then $u_i(\boldsymbol{o}^t) \geq u_i(\boldsymbol{o}^{t-1}) + \boldsymbol{x}^{t-1}_i \geq u_i(\boldsymbol{o}^{t-1}) + {qProp^1} +  \mathbbm{1}_{i \in {sN}^{t-1}} + \mathbbm{1}_{i \in \overline{{sQ}^{t-1}_>}}$.
However, since for any agent $i$, $u_i(\boldsymbol{o}^{t-1}) \geq qProp^{t-1} + \mathbbm{1}_{i \in {Q}^{t-1}_>}$ (conditions (1) and (2)), we get that $u_i(\boldsymbol{o}^t) \geq qProp^{t-1} + {qProp^1} + \mathbbm{1}_{i \in {Q}^{t-1}_>} + \mathbbm{1}_{i \in {sN}^{t-1}} + \mathbbm{1}_{i \in \overline{{sQ}^{t-1}_>}}$.

First, let us find a lower bound for $|{Q}^{t-1}_> \cup \overline{{sQ}^{t-1}_>}|$. Since $\overline{{sQ}^{t-1}_>} \subseteq \overline{{Q}^{t-1}_>}$ then ${Q}^{t-1}_> \cap \overline{{sQ}^{t-1}_>} =\emptyset$. We thus get that $|{Q}^{t-1}_> \cup \overline{{sQ}^{t-1}_>}| = |{Q}^{t-1}_>| + |\overline{{sQ}^{t-1}_>}| = |{Q}^{t-1}_>| + \min{(|\overline{{Q}^{t-1}_>}|, {rProp^1})} = |{Q}^{t-1}_>| + \min{(n - |{Q}^{t-1}_>|, {rProp^1})} =  \min{(n, {rProp^1} + |{Q}^{t-1}_>|)}$. Since condition (2) holds,     $\min{(n, {rProp^1} + |{Q}^{t-1}_>|)} \geq \min{(n, {rProp^1} + {rProp^{t-1})}}$. Overall,  $|{Q}^{t-1}_> \cup \overline{{sQ}^{t-1}_>}|  \geq  \min{(n, {rProp^1} + {rProp^{t-1})}}$.

Let $a = (t - 1) \cdot (m-1)$ and $b = (m-1)$. Then, $a + b = t \cdot (m-1)$.
Since 
\begin{equation} 
qProp^{t-1} = \floor{\frac{a}{n}}, qProp^{1} = \floor{\frac{b}{n}}, qProp^{t} = \floor{\frac{a + b}{n}},    
\end{equation}
and
\begin{equation}
    \begin{aligned}
       rProp^{t-1}  = a \pmod n, rProp^{1} = b \pmod n,\\
       rProp^{t}    = (a + b) \pmod n, 
    \end{aligned}
\end{equation}

then according to 
\autoref{lemma:postive_integer_properties}, 
\begin{equation}
    \begin{aligned}
        \label{eq:x=0}
        qProp^{t-1} + {qProp^1} = qProp^{t} \text{  and  }\\
        rProp^{t-1} + {rProp^1} = rProp^{t},   
   \end{aligned}
\end{equation}
or 
\begin{equation}
    \begin{aligned}
        \label{eq:x=1}
        qProp^{t-1} + {qProp^1} + 1 = qProp^{t} \text{  and }\\
        rProp^{t-1} + {rProp^1} = n + rProp^{t}.   
    \end{aligned}
\end{equation}

If  \autoref{eq:x=0} holds, then $u_i(\boldsymbol{o}^t) \geq qProp^{t-1} + {qProp^1} + \mathbbm{1}_{i \in {Q}^{t-1}_>} + \mathbbm{1}_{i \in {sN}^{t-1}} + \mathbbm{1}_{i \in \overline{{sQ}^{t-1}_>}} = qProp^{t} + \mathbbm{1}_{i \in {Q}^{t-1}_>} + \mathbbm{1}_{i \in \overline{{sQ}^{t-1}_>}} + \mathbbm{1}_{i \in {sN}^{t-1}}$. That is, $Q^t = C$ and $\{{Q}^{t-1}_> \cup \overline{{sQ}^{t-1}_>}\} \subseteq {Q}^{t}_>$. Thus, $|{Q}^{t}_>|  \geq |{Q}^{t-1}_> \cup \overline{{sQ}^{t-1}_>}| \geq \min{(n, {rProp^1} + {rProp^{t-1})}} = \min{(n, rProp^{t})} = rProp^{t}$. Overall, claims (3) and (4) hold.

On the other hand, if \autoref{eq:x=1} holds, then $u_i(\boldsymbol{o}^t) \geq qProp^{t-1} + {qProp^1} + \mathbbm{1}_{i \in {Q}^{t-1}_>} + \mathbbm{1}_{i \in {sN}^{t-1}} + \mathbbm{1}_{i \in \overline{{sQ}^{t-1}_>}} = qProp^{t} - 1 + \mathbbm{1}_{i \in {Q}^{t-1}_>} + \mathbbm{1}_{i \in \overline{{sQ}^{t-1}_>}} + \mathbbm{1}_{i \in {sN}^{t-1}}$. Note that for every agent $i$ such that $i \in {Q}^{t-1}_>$ or $i \in \overline{{sQ}^{t-1}_>}$, $u_i(\boldsymbol{o}^t) \geq qProp^{t} + \mathbbm{1}_{i \in {sN}^{t-1}}$. That is, for any agent $i$ such that $i \in {Q}^{t-1}_>$ or $i \in \overline{{sQ}^{t-1}_>}$, $i \in Q^t$. We thus get that $\{{Q}^{t-1}_> \cup \overline{{sQ}^{t-1}_>}\} \subseteq {Q}^t$. That is, $|{Q}^t| \geq |{Q}^{t-1}_> \cup \overline{{sQ}^{t-1}_>}| \geq \min{(n, {rProp^1} + {rProp^{t-1})}} = \min{(n, n + rProp^{t})} = n$. Therefore, ${Q}^t = C$. That is, claim (3) holds. Moreover, we can now state that for any agent $i$, $u_i(\boldsymbol{o}^t) \geq qProp^{t} + \mathbbm{1}_{i \in {sN}^{t-1}}$, and thus ${sN}^{t-1} \subseteq {Q}^{t}_>$.

We now find $|{Q}^{t}_>|$. 
Since ${sN}^{t-1} \subseteq {Q}^{t}_>$,
then $|{Q}^{t}_>| \geq |{sN}^{t-1}|$. By definition, $ |{sN}^{t-1}| = {rProp^1} - |\overline{{sQ}^{t-1}_>}| = {rProp^1} - \min{(|\overline{{Q}^{t-1}_>}|, {rProp^1})}$.
That is, $\min{(|\overline{{Q}^{t-1}_>}|, {rProp^1})} + |{Q}^{t}_>| \geq {rProp^1}$.
Since $|\overline{{Q}^{t-1}_>}| = n -  |{Q}^{t-1}_>|$ and $|{Q}^{t-1}_>| \geq {rProp}^{t-1}$ then $|\overline{{Q}^{t-1}_>}| \leq n - {rProp}^{t-1}$.
Therefore, $\min{(n - {rProp}^{t-1}, {rProp^1})} + |{Q}^{t}_>| \geq \min{(|\overline{{Q}^{t-1}_>}|, {rProp^1})} + |{Q}^{t}_>| \geq {rProp^1}$.
According to \autoref{eq:x=1},  $rProp^{t-1} + {rProp^1} = n + rProp^{t}$, and thus $\min{({rProp^1} - rProp^{t}, {rProp^1})} + |{Q}^{t}_>| \geq  {rProp^1}$. That is, ${rProp^1}  - rProp^{t} + |{Q}^{t}_>| \geq  rProp^{1}$. We thus get that  $- rProp^{t} + |{Q}^{t}_>| \geq  0$. That is, $|{Q}^{t}_>| \geq rProp^{t}$. That is, claim (4) holds.

Overall, we get that if conditions (1) and (2) hold, then there exists a candidate $c$ such that  claims (3) and (4) hold.
\end{proof}

We are now ready to prove the approximation result.
\begin{theorem}
\label{theorem:1-additive-prop-leximin}
$\textsf{LMin}_{\textsf{on}}$ guarantees a $1$-additive approximation of $Prop$.
\end{theorem}
\begin{proof}
Let $\boldsymbol{o}^{t-1}$ be the vector of chosen candidates by $\textsf{LMin}_{\textsf{on}}$ up to round $t$. If $\boldsymbol{o}^{t-1}$ satisfies conditions (1) and (2) of \autoref{lemma:lemma_addtive_main}, then there exists a candidate $c$ such that conditions (3) and (4) of \autoref{lemma:lemma_addtive_main} hold. By definition, $\textsf{LMin}_{\textsf{on}}$ chooses a candidate $o^t$ such that 
 $u(\boldsymbol{o}^t) \succeq u((o^1,\ldots,o^{t-1},c))$. That is, $\boldsymbol{o}^t$ that is chosen by $\textsf{LMin}_{\textsf{on}}$ satisfies conditions (3) and (4). In addition, in the first round $t = 1$ and thus, obviously, $rProp^{t-1} = qProp^{t-1} = 0$. That is, conditions (1) and (2) of \autoref{lemma:lemma_addtive_main} hold when $t=1$.
Therefore, for any round $t$ and agent $i$, $u_i(\boldsymbol{o}^{t}) \geq qProp^t$. Specifically, when $t=T$, $u_i(\boldsymbol{o}^{T}) \geq qProp^T = \floor{Prop_i}$. By definition, $\floor{Prop_i} > Prop_i - 1$, and $\boldsymbol{o} =\boldsymbol{o}^{T}$.
That is, $u_i(\boldsymbol{o}) + 1 > Prop_i$, for any agent $i$.
\end{proof}

Similar to the offline setting, the other mechanisms do not guarantee an additive constant-factor approximation of $Prop$. Indeed, $\textsf{RR}_{\textsf{on}}$ is essentially equivalent to $\textsf{RR}_{\textsf{off}}$ when the valuations are restricted to Borda valuations, since for every agent $i$ and in every round $t$, 
$cMax^t_i$ are the same (i.e., $m-1$).
That is, $\textsf{RR}_{\textsf{on}}$ does not guarantee an additive approximation that is better than $m-1$.
As for $\textsf{MNW}_{\textsf{on}}$, since the example in the proof of \autoref{theorem:mnw_not_m-5_app} hold for any $t \leq T$, $\textsf{MNW}_{\textsf{on}}$ chooses the same outcome as $\textsf{MNW}_{\textsf{off}}$ in this instance. That is, $\textsf{MNW}_{\textsf{on}}$ does not guarantee an additive approximation of $Prop$ that is better than $\frac{m-3}{2}$.

We now analyze $MPP$. 
Unfortunately, $MPP$ cannot be satisfied in the online setting, even with Borda valuations.

\begin{theoremrep}
Even with Borda valuations, there is no online mechanism that satisfies $MPP$.
\end{theoremrep}
\begin{proof}
Assume by contradiction that there is an online mechanism that satisfies $MPP$. Let $m = 2$, $n = 4$, $T = 2$, and
$$
V^1  =  \begin{pmatrix}
1 & 0\\
1 & 0\\
0 & 1\\
0 & 1
\end{pmatrix},  X = \begin{pmatrix}
1 & 0\\
0 & 1\\
0 & 1\\
0 & 1
\end{pmatrix}, Y = \begin{pmatrix}
1 & 0\\
1 & 0\\
1 & 0\\
0 & 1
\end{pmatrix}.
$$
If $V^2 = X$, the only outcome that satisfies $Prop$ is $(c_1,c_2)$. Therefore, the mechanism must choose the candidate $c_1$ in the first round.
On the other hand, if $V^2 = Y$, the only outcome that satisfies $Prop$ is $(c_2,c_1)$, and thus the mechanism must choose $c_2$ in the first round, which is a contradiction.
\end{proof}


Next, we analyze $RRS$.
\begin{theoremrep}
With borda valuations, $\textsf{LMin}_{\textsf{on}}$ satisfies $RRS$.     
\end{theoremrep}
 \begin{proof}
Let $\boldsymbol{o}$ be the outcome chosen by $\textsf{LMin}_{\textsf{on}}$.
Recall that for any agent $i$, $RRS_i =\floor{\frac{T}{n}} \cdot (m-1)$ with Borda valuations. According to the proof of ~\autoref{theorem:1-additive-prop-leximin}, for any agent $i$, $u_i(\boldsymbol{o}) \geq qProp^T = \floor{\frac{T\cdot (m-1)}{n}}$. Clearly, $\floor{\frac{T}{n}(m-1)} \geq \floor{\frac{T}{n}} \cdot (m-1)$. We thus get that for any agent $i$, $u_i(\boldsymbol{o}) \geq RRS_i$. That is, $\textsf{LMin}_{\textsf{on}}$ satisfies $RRS$.  
\end{proof}

As for $\textsf{MNW}_{\textsf{on}}$, since the example in the proof of \autoref{theorem:mnw_rr_not_rrs_mpp} hold for $t \leq T$, $\textsf{MNW}_{\textsf{on}}$ chooses the same outcome as $\textsf{MNW}_{\textsf{off}}$ in this instance. That is, $\textsf{MNW}_{\textsf{on}}$ does not satisfy $RRS$.

Finally, we consider $PO$.
Recall that $\textsf{LMin}_{\textsf{off}}$ and $\textsf{MNW}_{\textsf{off}}$ satisfy $PO$. In the online setting, even with Borda valuations, they do not satisfy $PO$ anymore.
\begin{theoremrep}
Even with Borda valuations, $\textsf{LMin}_{\textsf{on}}$ and  $\textsf{MNW}_{\textsf{on}}$ do not satisfy $PO$.
\end{theoremrep}
\begin{proof}
Let $m = 9, n = 2, T = 2$, and 
$$
V^1 = V^2 =  \begin{pmatrix}
8 & 7 & 6 & 5 & 4 & 3 & 2 & 0 & 1\\
1 & 0 & 2 & 3 & 4 & 5 & 6 & 7 & 8
\end{pmatrix}.
$$
$\textsf{LMin}_{\textsf{on}}$ and $\textsf{MNW}_{\textsf{on}}$ choose $\boldsymbol{o} = (c_5,c_5)$, while the $PO$ outcomes are $(c_1,c_9)$ or $(c_9,c_1)$.
\end{proof}

\begin{toappendix}
\section{Efficiency and Fairness for Online Setting}
\FloatBarrier
\begin{table}[ht]
    \centering
\setlength{\extrarowheight}{1pt}
\setlength\arrayrulewidth{2pt}
\begin{tabular}{@{\extracolsep{\fill}} |c|c|c|c|c|c}
 \hline
  & Local $PO$ & IAT & Approval & Homogeneity\\
\hline
$\textsf{LMin}_{\textsf{on}}$ & \cmark & \cmark & \cmark & \cmark\\
\hline
$\textsf{RR}_{\textsf{on}}$ & \xmark & \cmark & \xmark  & \xmark\\
 \hline
$\textsf{MNW}_{\textsf{on}}$ & \cmark & \xmark & \cmark  & \cmark\\
\hline
\end{tabular}
 \caption{ Summary of the results for the online setting, where there are no restrictions on the valuations.}
 \label{tab:axioms}
\end{table}
Recall that with no restriction on the valuations, all the mechanisms that we study do not satisfy the $PO$ efficiency property in the online setting. In addition, in \autoref{sec:online} we showed that it is impossible to satisfy $Prop1$ or $\alpha$-$RRS$ for any constant $\alpha$.
We thus propose weaker notions of efficiency and fairness.
Specifically, for efficiency, we propose the notion of local Pareto optimality, which is defined as follows:
\begin{definition}
An outcome $\boldsymbol{o}$ is \textit{local Pareto Optimal} (local $PO$) if for every round $t$, there does not exist a candidate $c$ such that for every agent $i$, (1) $v^t_i(c) \geq v^t_i(o^t)$ and (2) there exists an agent $j$ such that $v^t_j(c) > v^t_j(o^t)$.
\end{definition}
Clearly, $\textsf{LMin}_{\textsf{on}}$ and $\textsf{MNW}_{\textsf{on}}$ satisfy local $PO$, but $\textsf{RR}_{\textsf{on}}$ does not satisfy local $PO$.

For fairness, we propose the Invariant to Affine Transformations ($IAT$) axiom, which is adapted from bargaining games \cite{john1953nash}. 

\begin{definition}
Given any $\alpha \in \mathbb{R^+}$ and $\beta \in \mathbb{R}$ such that all the valuations $\overline{v^t_i(c_j)} = v^t_i(c_j) \cdot \alpha + \beta$ are non-negative, let $\overline{\boldsymbol{o}}$ be the outcome of the mechanism when the valuations are $\overline{v^t_i(c_j)}$. A mechanism satisfies Invariant to Affine Transformations (IAT) if $u(\boldsymbol{o}) = u(\overline{\boldsymbol{o}})$.
\end{definition}
Clearly, $\textsf{RR}_{\textsf{on}}$ satisfy $IAT$, since $cMax^t_i$ remains the same candidate even after an affine transformation of the valuations. $\textsf{LMin}_{\textsf{on}}$ also satisfy $IAT$ since if $\boldsymbol{x} \succ \boldsymbol{y}$ then $\alpha \cdot \boldsymbol{x} + \beta \cdot \boldsymbol{1}_{n} \succ \alpha \cdot \boldsymbol{y} + \beta \cdot \boldsymbol{1}_{n}$.
However, $\textsf{MNW}_{\textsf{on}}$ does not satisfy $IAT$.
\begin{theoremrep}
 $\textsf{MNW}_{\textsf{on}}$ does not satisfy $IAT$. 
\end{theoremrep}
\begin{proof}
Let $m = 2, n = 2,T = 1$, and
$$
V^1 =  \begin{pmatrix}
5 & 1\\
0 & 1 
\end{pmatrix}.
$$ $\textsf{MNW}_{\textsf{on}}$ chooses $\boldsymbol{o} = (c_2)$. However, if   $\overline{v^t_i(c_j)} = v^t_i(c_j) + 1$, then $\textsf{MNW}_{\textsf{on}}$ chooses $\overline{\boldsymbol{o}} = (c_1)$.
\end{proof}

Next, we propose $approval$, which is a stronger variant of the Plurality axiom that was proposed by \cite{freeman2017fair}. Intuitively, a mechanism satisfies $approval$ if whenever all agents' valuations are dichotomous- they assign either $1$ or $0$ to each candidate, and they all have the same accumulated utilities, then the candidate with the highest number of $1$'s is chosen. Formally,
\begin{definition}
A mechanism satisfies $approval$ if for a given $t$, for every agents $i,j \in N$, ${u_i(\boldsymbol{o}^{t-1})} = {u_j(\boldsymbol{o}^{t-1})}$, and for every agent $i \in N$ and candidate $c \in C$, $v^t_i(c) \in \{0,1\}$, then
$o^t \in \argmax_{c \in C}{\sum_{i \in N} v^t_{i}(c)}$.
\end{definition}
Clearly, $\textsf{RR}_{\textsf{on}}$ does not satisfy $approval$. However, by definition, $\textsf{LMin}_{\textsf{on}}$ and $\textsf{MNW}_{\textsf{on}}$ satisfy $approval$. 

Finally, we consider the axiom of homogeneity, which requires that if every agent's ballot is replicated the same number of times, the outcome remains the same (or it is replaced with an outcome with the same utility). Formally,
\begin{definition}
Given a round $t$, let $x$ be a positive integer. For any $k \in \mathbb{N}, 1 \leq k \leq t$, let ${V'}^k$ be a block matrix that consists of $x$ copies of $V^k$. That is, ${V'}^k =
\begin{bmatrix}
V^k\\
\vdots\\
V^k\\
\end{bmatrix}$.
Let ${o'}^t$ be the outcome of the mechanism at round $t$, when the input is $({V'}^1, \ldots, {V'}^t)$ and the vector of chosen candidates is $\boldsymbol{o}^{t - 1}$.
A mechanism satisfies homogeneity if $\boldsymbol{u}(o^t) = \boldsymbol{u}({o'}^t)$.
\end{definition}
Clearly, $\textsf{RR}_{\textsf{on}}$ does not satisfy $homogeneity$ for all the orders $\pi$. However, $\textsf{LMin}_{\textsf{on}}$ satisfies $homogeneity$ since if $\boldsymbol{x} \succ \boldsymbol{y}$ then $(x, \ldots, x) \succ (y, \ldots, y)$. In addition, $\textsf{MNW}_{\textsf{on}}$ satisfies $homogeneity$ since $\argmax_{c_j \in C} \prod_{i \in N} (u_i(\boldsymbol{o}^{t-1}) + v^t_i(c_j)) = \argmax_{c_j \in C} \prod_{i \in N} (u_i(\boldsymbol{o}^{t-1}) + v^t_i(c_j))^{x}$.
\end{toappendix}
\section{Conclusions and Future Work}
In this paper, we study collective decision-making mechanisms for a sequence of decisions, when the preferences of the agents are cardinal. However, since we also concentrate on the restriction to Borda valuations, our results are also applicable when the preferences of the agents are ordinal.
Overall, we claim that the leximin mechanism, in which the valuations are normalized with the proportional value of each agent, has a significant advantage over the other known mechanisms: in the offline setting, it satisfies $PO$ and $MPP$, and when there are only two agents it also satisfies $1/2$-$Prop1$ and $RRS$. With Borda valuations, the leximin mechanism satisfies $Prop1$ and $RRS$, and guarantees the best possible constant-factor approximation of $Prop$. Moreover, these results hold both for the offline and online settings.

For future work, we would like to consider other restrictions on the agents' valuations, e.g., valuations that are based on a (given) general scoring rule. We would also like to study what happens when there are many rounds. Specifically, we conjecture that given any $n$ and $m$, there is a $t$ such that the leximin mechanism with Borda valuations satisfies $Prop$ for any $T \geq t$. Finally, since it is impossible to achieve proportionality or its relaxations in the online setting, we would like to extend our framework and analyze probabilistic mechanisms. 

\section*{Acknowledgments}
This research has been partly supported by the Ministry of Science, Technology \& Space (MOST), Israel.


\bibliographystyle{plainnat}
\bibliography{comsoc2023}


\begin{contact}
Ido Kahana\\
Ariel University\\
Ariel, Israel\\
\email{ynet.ido@gmail.com}
\end{contact}

\begin{contact}
Noam Hazon\\
Ariel University\\
Ariel, Israel\\
\email{noamh@ariel.ac.il}
\end{contact}


\end{document}